\documentclass[fleqn,usenatbib]{mnras}
\usepackage{newtxtext,newtxmath}
\usepackage[T1]{fontenc}
\usepackage{graphicx}	
\usepackage{amsmath}	
\usepackage{enumitem}
\usepackage[english]{babel}
\usepackage{gensymb}
\usepackage{hyperref}
\usepackage{comment}
\usepackage{xcolor}
\definecolor{darkblue}{rgb}{0.0, 0.0, 0.55}
\definecolor{darkred}{rgb}{0.55, 0.0, 0}
\definecolor{gray}{rgb}{0.4, 0.4, 0.4}
\definecolor{darkgray}{rgb}{0.2, 0.2, 0.2}

\title[Primary beam effects on foreground cleaning]{\textsc{Hi} intensity mapping with MeerKAT: Primary beam effects on foreground cleaning}

\author[S.D. Matshawule et al.]{Siyambonga D. Matshawule$^{1}$  \thanks{E-mail: smatshawule@uwc.ac.za}, Marta Spinelli$^{2,3,1}$, Mario G. Santos$^{1,4}$, Sibonelo Ngobese$^{1}$
\\
$^{1}$ Department of Physics and Astronomy, University of the Western Cape, Robert Sobukhwe Road, Bellville, 7535, South Africa\\
$^{2}$INAF-Osservatorio Astronomico di Trieste, Via G.B. Tiepolo 11, 34143 Trieste, Italy\\
$^{3}$ IFPU - Institute for Fundamental Physics of the Universe, Via Beirut 2, 34014 Trieste, Italy\\
$^{4}$ South African Radio Observatory (SARAO), 2 Fir Street, Observatory, Cape Town, 7925, South Africa
}

\date{Accepted XXX. Received YYY; in original form ZZZ}

\pubyear{2020}

\begin{document}
\label{firstpage}
\pagerange{\pageref{firstpage}--\pageref{lastpage}}
\maketitle
\begin{abstract}
Upcoming and future neutral hydrogen Intensity Mapping surveys offer a great opportunity to constrain cosmology in the post-reionization Universe, provided a good accuracy is achieved in the separation between the strong foregrounds and the cosmological signal. Cleaning methods are often applied under the assumption of a simplistic Gaussian primary beam. In this work, we test the cleaning in the presence of a realistic primary beam model with a non-trivial frequency dependence. We focus on the Square Kilometre Array precursor MeerKAT telescope and simulate a single-dish wide-area survey. We consider the main foreground components, including an accurate full-sky point source catalogue. We find that the coupling between beam sidelobes and the foreground structure can complicate the cleaning. However, when the beam frequency dependence is smooth, we show that the cleaning is only problematic if the far sidelobes are unexpectedly large. Even in that case, a proper  reconstruction is possible if the strongest point sources are removed and the cleaning is more aggressive.
We then consider a non-trivial frequency dependence: a sinusoidal type feature in the beam width that is present in the MeerKAT beam and is expected in most dishes, including SKA1-MID. Such a feature, coupling with the foreground emission, biases the reconstruction of the signal across frequency, potentially impacting the cosmological analysis. We show that this effect is constrained to a narrow region in $k_\parallel$ space and can be reduced if the maps are carefully re-smoothed to a common lower resolution.
\end{abstract}




\section{Introduction}

The past few decades have seen major advances in observational cosmology. In this era of "precision cosmology" most of the data has come from two sources: the cosmic microwave background (CMB) experiments \citep[e.g.][]{planck2018} providing a two dimensional view of the early Universe ($z\sim 1000$), and observations at optical wavelengths, such as supernova surveys (\citealt{riess1998}; \citealt{perlmutter1999}) and observations of the three dimensional large scale structure using galaxy redshift surveys, e.g. \citealt{anderson2014}; \citealt{contreras2013}, allowing to probe the late-time Universe ($z<1$).  

As the push towards greater cosmological precision continues, it becomes necessary to survey progressively larger volumes of the Universe in order to beat cosmic variance. Large galaxy spectroscopic surveys are planned for this effect (DESI, Euclid). In the radio, the 21cm line provides a straightforward, three dimensional, way to trace the neutral hydrogen (HI) which is mostly immune to obscuration by intervening matter. Galaxy spectroscopic surveys are however quite time consuming, which limits their capacity to probe large volumes and wide redshift ranges. HI Intensity Mapping (HI IM), has been proposed as a way to circumvent this problem \citep{battye2004,mcquinn2006,chang2008,mao2008,loebwyithe2008, pritchard2008,wyitheloeb2008,wyithe2008,peterson2009, bagla2010, seo2010, lidz2011,ansari2012,battye2013}. 
HI IM relies on measuring the total radiation intensity across the sky and as a function of redshift, and can quickly cover large volumes, although with low angular resolution. It also provides high redshift resolution given the relation between observed radio frequency and redshift for the 21cm line.

Only a few HI IM surveys have been done so far. The HI IM technique was first tested with the Green Bank Telescope (GBT), by measuring the cross-correlation function between HI IM and optical galaxies \citep{chang2010}, which was further improved in \citet{masui2013,li2014,wolz2017}. Cross correlation measurements with the Parkes telescope were also done, although at smaller cosmological scales \citep{anderson2018}, while the HI IM auto power spectrum remains undetected \citep{switzer2013}.
Several HI IM experiments are being planned (see \citealt{kovetz2017}), such as the Tianlai project \citep{chen2012}, the
Canadian Hydrogen Intensity Mapping Experiment (CHIME \cite{bandura2014}), the
Baryonic Acoustic Oscillations from Integrated Neutral Gas Observations 
(BINGO \cite{battye2013}) and the
Hydrogen Intensity and Real-Time Analysis experiment (HIRAX \cite{newburgh2016}). 
The SKA has been proposed as a major instrument to probe cosmology using this technique using the single dish information \citep{santos2015, bull2015, skawg2020}. More recently, it was also proposed to perform an HI IM survey with the newly built MeerKAT telescope in single-dish mode \citep{santos2017}.

The success of the HI IM technique relies on being able to separate the cosmological HI signal from the strong foreground
radio sources emitting in the same frequency range. The strongest of these, synchrotron emission from our own galaxy, is about 5 orders of magnitude larger than the expected 21cm signal, even at high galactic latitudes. Likewise, extra-galactic point sources can be about 3 orders of magnitude stronger than the 21cm signal (\citealt{dimatteo2002}; \citealt{ohmack2003}; \citealt{santos2005}). Fortunately, as opposed to the cosmological signal, most relevant foregrounds have a very smooth frequency dependence or other statistical properties that can be exploited to subtract them efficiently (\citealt{lui2009}; \cite{lui2011}; \citealt{masui2013}; \citealt{wolz2014}; \citealt{shaw2014,shaw2015}; \citealt{alonso2014}). 
Nevertheless, their large amplitude can still create serious problems if the cleaning is not done with extremely high accuracy. Interloper lines from higher redshifts (e.g. OH) and radio recombination lines from our Galaxy at the target frequencies (e.g. H187$\alpha$) should be much smaller than the HI signal \citep{gong2011, battye2013}.

One of the main problems with foreground removal is that the instrument itself can complicate the foreground emission or even add extra systematics. 
Examples include amplitude gain fluctuations (e.g. \citealt{li2020}), mixing of polarization components or even digital non-linearities. Other possibilities are internal signal chain reflections and antenna cross-coupling which have been studied in the context of compact low-frequency arrays (e.g. \citealt{kern2019}). The solution relies on a careful calibration of the telescope and understanding such systematics. For example, one of the dominant effects, the amplitude gain frequency fluctuations seen in the autocorrelations, can be corrected by observing a few strong point sources and comparing their expected smooth spectra with the observed ones (the standard bandpass calibration). With foregrounds up to $10^5$ times stronger than the 21cm signal this requires high accuracy and an instrument that is stable enough in time.

The primary beam can also change the frequency structure of the foregrounds due to its own frequency dependence. This can be particularly insidious for point sources "sitting" on the sidelobes of the beam where changes happen faster. In principle, knowing the shape of the beam at all frequencies, either from calibration or previous measurements, would allow to correct for such effect. This is however quite difficult to achieve, in particular far away from the beam center. It is also complicated by the fact that the beam can change between different dishes in the array and over time.
Understanding how damaging such contributions can be to the overall detection of the signal is therefore of crucial importance. 
This problem has been studied in the context of interferometric observations at low frequencies to probe the Epoch of Reionization (EoR), using foreground avoidance techniques.
The natural chromaticity of an interferometer creates a well-defined region  in  Fourier  space  called  the foreground wedge \citep[e.g.][]{datta2010,vedantham2012,trott2012,pober2014} but the 
spectral structure of the foregrounds and the chromaticity of the antenna gains contribute to spread foreground power beyond the wedge into the EoR window,
as seen in the data \citep[e.g.][]{parson2012,pober2013,thyagarajan2016} and studied in simulations \citep[e.g.][]{lanman2020}.

This paper studies the effect of the primary beam frequency dependence on the foregrounds and its impact on the recovery of the HI IM signal, in single dish mode, for post-reionization frequencies.
We study foreground cleaning in particular in the presence of strong point sources. We focus on the MeerKAT telescope considering part of its L-band (900-1600~$\mathsf{MHz}$) and a survey exploiting the auto-correlation data, although most of the conclusions can be applied to other telescopes, such as the future SKA (in single dish observations). We assume that calibration has been done already (see \citealt{wang2020} for an example of the pipeline) and neglect any effects coming from the gains. We analyze the data convolved by our primary beam models and provide different tests of its impact on the signal extraction. The beams are assumed normalized to 1 at the center and therefore, any overall amplitude effect (possibly frequency dependent), is assumed to have already been absorbed in the bandpass calibration. Further effects will be considered as we expand our simulation pipeline.
The paper is organized as follows: in section~\ref{sec:beam} we present possible models of the MeerKAT beam; in section~\ref{sec:skymodel} we describe the signal and foreground simulations used in this work, presenting in particular a new point source mock catalogue; in section~\ref{sec:beamconv} we discuss how the sky model is convolved with the beam to mimic realistic observations; in section~\ref{sec:pipe} we summarize the simulated survey specifications and detail the simulation pipeline; section~\ref{sec:results} is dedicated to the discussion of our results; in section~\ref{sec:conclusions} we present our conclusions.

\section{MeerKAT beam model}\label{sec:beam}

\begin{figure}
    \includegraphics[width=9cm]{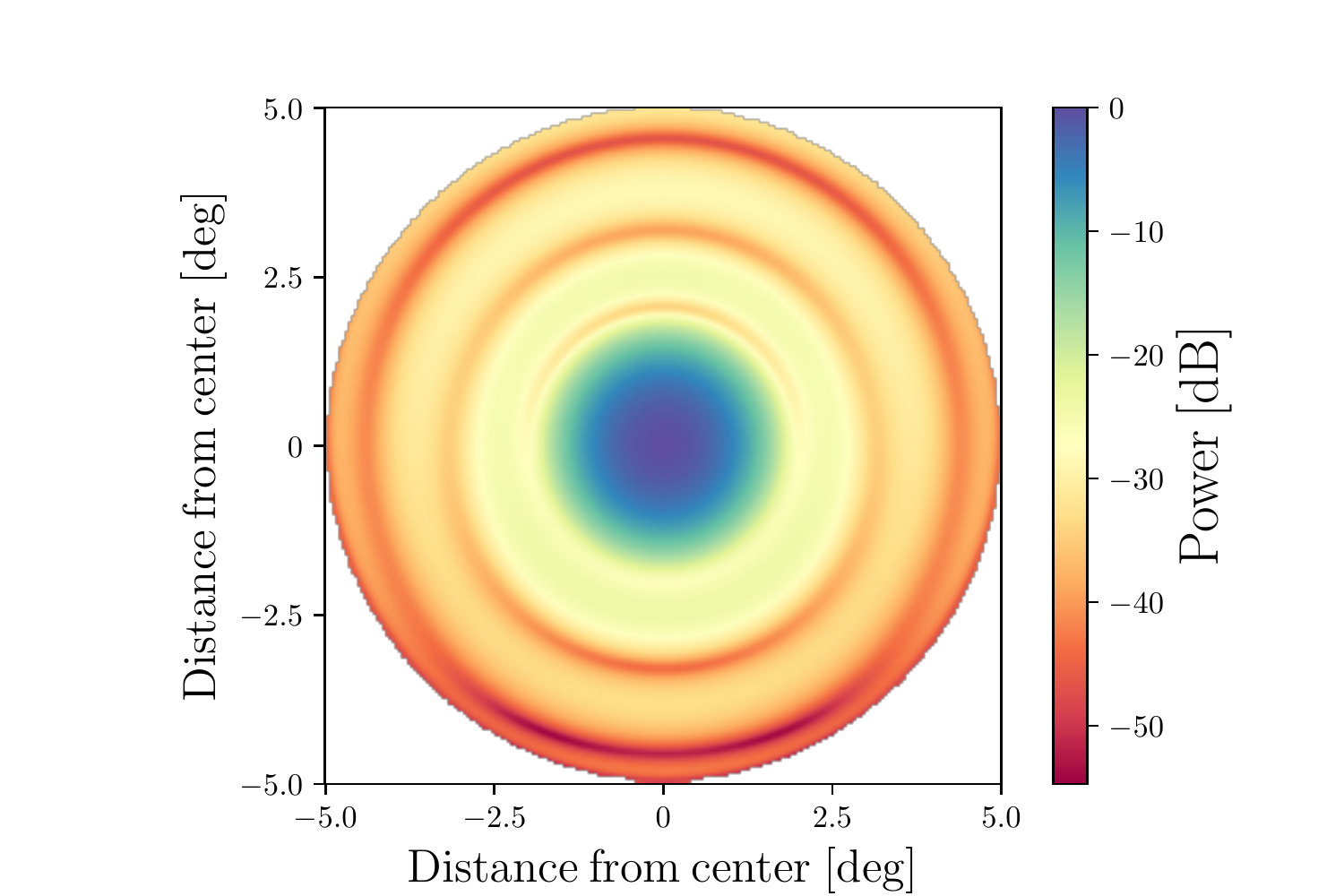}
    \caption{$2$-d image of the MeerKAT primary beam model at $950~\mathsf{MHz}$ obtained using the Eidos package \citep{asad2019}.} 
    \label{fig:2d_eidos_beam}
\end{figure}

We start by describing the telescope primary beam models used in this paper. We concentrate on the MeerKAT single dish observations (auto-correlation signal) and assume for simplicity that all dishes have the same primary beam, since our focus are the frequency effects. This is a fair approach since map making is done separately for every dish. We  only look at the total intensity (Stokes I) since the 21cm signal is unpolarized to a high degree. Polarization leakage might be an important effect for foreground cleaning \citep{alonso2014,carucci2020,cunnington2020b} and we plan to return to it in a follow up paper. A final assumption is that our beam model has circular symmetry. Indeed, as seen in figure~\ref{fig:2d_eidos_beam}, the MeerKAT beam is symmetric to a good degree. There are however small beam asymmetries that become more relevant away from the center \citep{asad2019}. Nevertheless, this is more of an issue when we need to "subtract" point sources from the map and is an effect that can be separated from the frequency problem which is our focus here. We can imagine that each pixel on the sky will be observed many times. The beam we are modelling will be the final one after averaging over these observations and should therefore be more symmetric. Deviations from this symmetry means that point sources along a given circumference might be multiplied by a different beam value. However, the changes in frequency and its impact on smoothness will still be similar in any given direction and we believe that the models presented here capture the most relevant effects. 

Using spherical coordinates, we represent our beam function as $B(\nu,\theta,\phi)$, where $\nu$ is the frequency of observation, $\theta$ the polar angle and $\phi$ the azimuthal angle. The beam pattern is related to the dish reflective surface, or more accurately, its effective area set by the aperture illumination function, through a Fourier transform (for details see \citealt{wilson2013}). It is maximal in the direction at which the telescope is pointing ($\theta=0$) and decreases with $\theta$, away from the pointing direction. As already mentioned, we assume that gain calibration has been done already and the beams are normalized to 1 at the center.

\begin{figure}
\includegraphics[width=9cm]{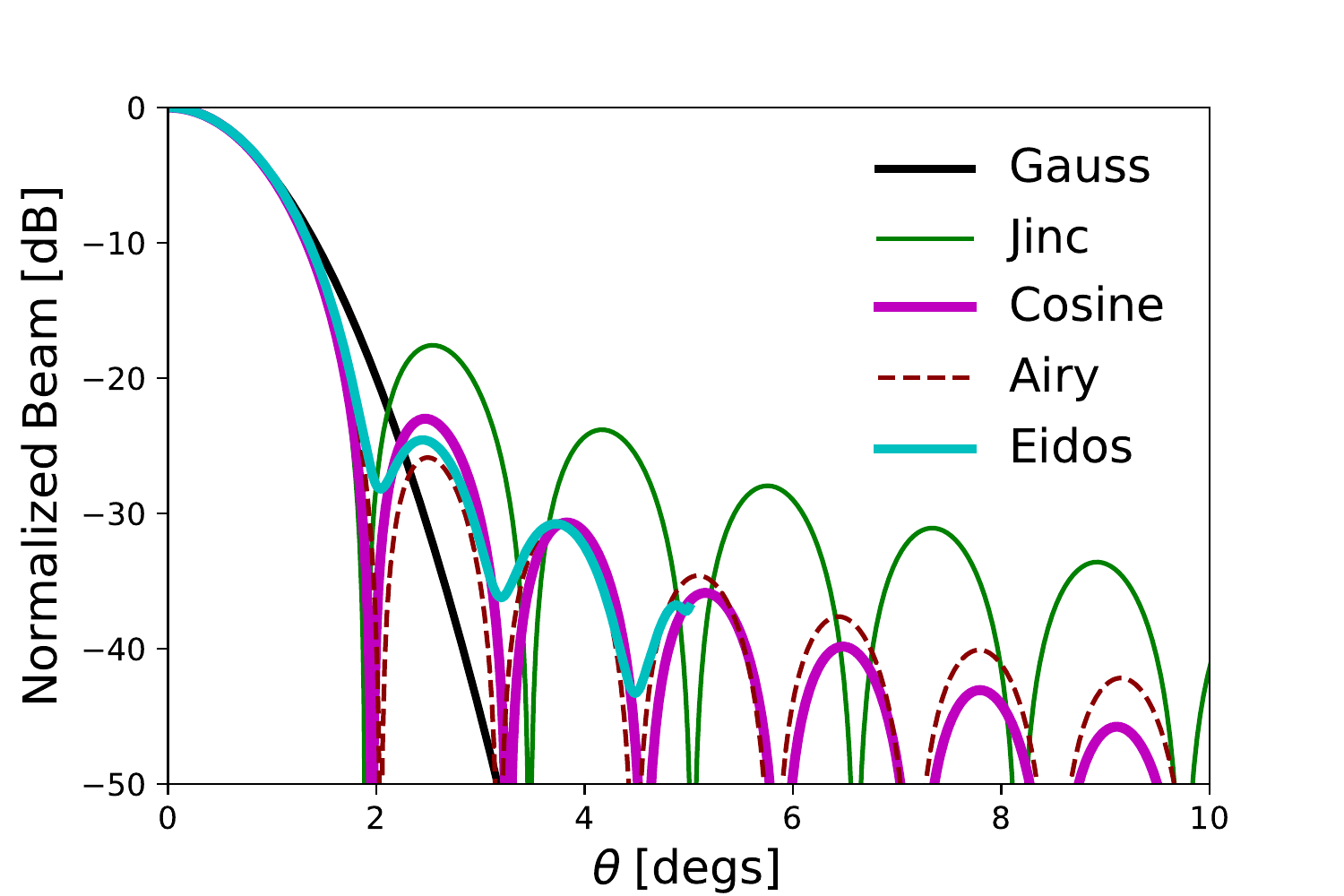}
\caption{A comparison of primary beam models at $950~\mathsf{MHz}$: the standard Gauss model (black), the Jinc beam model corresponding to an unblocked circular aperture \citep{wilson2013} (solid green), the cosine model \citep{condonransom2016} (magenta), the gaussian tapered airy disk used in \citet{harper2018} (dashed red) and the one obtained from the Eidos package presented in \citet{asad2019} (cyan).}
\label{fig:beam_models}
\end{figure}

Figure \ref{fig:2d_eidos_beam} shows the 2-dimensional MeerKAT beam at $950~\mathsf{MHz}$, using the "Eidos" package\footnote{\href{https://github.com/ratt-ru/eidos}{https://github.com/ratt-ru/eidos}} from \citet{asad2019}. The current version of the package can be used to create MeerKAT L-band beams from Zernike polynomial fits to holographic observations or EM simulations within a maximum diameter of 10 degrees. In this paper we use the fit to the MeerKAT holographic measurements. Using this package, we can generate images of the Stokes I beam at any frequency, although only up to 5 degrees from the beam center. We then numerically average this 2d beam over the $\phi$ direction in order to obtain a beam pattern that is a function of $\theta$ only: $B(\nu,\theta)$. We refer to this beam as the {\it Eidos} beam. In figure \ref{fig:beam_models} we can see the shape of this beam as a function of $\theta$, at a frequency of $\nu=950~~\mathsf{MHz}$ (solid cyan line).

Since the beam above only extends to 5 degrees, we explored other models that can be applied to full sky. High accuracy is not required but we want a function that captures some of the main trends of the MeerKAT beam: 1) it is an accurate representation within the main lobe; 2) has a full width at half maximum (FWHM) with the same frequency dependence; 3) decays with $\theta$ at the same rate and 4) has sidelobes and follows the nulls and peaks with reasonable accuracy. We would also like such function to be easy to calculate in order to quickly deploy it in simulations. 

\begin{figure}
  \includegraphics[width=9cm]{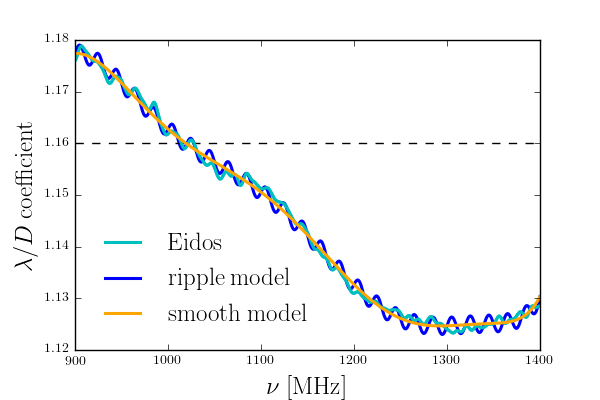}
   \caption{The FWHM of the azimuthally averaged MeerKAT/Eidos beam normalized by $\lambda/D$ as a function of frequency (solid cyan line) compared to our ripple model (solid blue line) which is composed of a sinusoidal oscillation on top of a smooth polynomial frequency dependence (solid orange line) - see equation~\ref{eq:fwhm_nu}.}
  \label{fig:fwhmVsfreq}
\end{figure}

The FWHM, $\Delta\theta$, of a MeerKAT dish is given approximately by
\begin{equation}
 \Delta\theta \approx 1.16\frac{\lambda}{D},\ 
 \label{eq:fwhm}
\end{equation}
where $\lambda$ is the observed wavelength and $D$ the dish diameter (13.5m for MeerKAT). 
Measurements of the MeerKAT/Eidos beam shows that the FWHM follows this dependence but also exhibits a low-level frequency-dependent {\it ripple}. This effect can be seen clearly in figure~\ref{fig:fwhmVsfreq}, where the Eidos FWHM is normalized by $\lambda/D$. This ripple is caused by the interaction between the primary and secondary reflector of MeerKAT \citep{villiers2013} and can be important in the foreground cleaning as it will add extra structure to the frequency spectra. 

\begin{table}
\centering
  \caption{Numerical values of the coefficient of equation~\ref{eq:fwhm_nu}.}
  \begin{tabular}{c c c }
  \hline
 A [arc-min] & T [MHz]  & $a_n$ $\{n=0,...,8\}$ \\
  \hline
 & & $\{6.7\mathrm{e}{3}, -50.3, 0.16,$\\ 
 0.1  & 20 & $-3.0\mathrm{e}{-4},3.5\mathrm{e}{-7}, -2.6\mathrm{e}{-10},$\\
 & & $1.2\mathrm{e}{-13}, -3.0\mathrm{e}{-17}, 3.4\mathrm{e}{-21} \}$ \\
  \hline
  \label{tab:ripple}
\end{tabular}
\end{table}

To address the effect on the extraction of the cosmological signal of such a frequency dependent FWHM, we model its main features.
We fit the additional smooth frequency dependence with a polynomial. We choose a high order polynomial (8th degree) to accurately describe the beam in the frequency interval of interest. On top of it we superimpose a sinusoidal oscillation with period $T$ and amplitude $A$ arc-minutes, 
\begin{equation}
      \centering
     \Delta\theta_r = \frac{\lambda}{D}\left(\sum_{d=0}^{8}a_d \nu^{d} + A\sin\Bigg(\frac{2\pi\nu}{T}\Bigg)\right).
     \label{eq:fwhm_nu}
\end{equation}
The values of the parameters are summarized in table~\ref{tab:ripple} and the ripple model and its smooth component are shown in figure~\ref{fig:fwhmVsfreq}. 
We note that in \citet{asad2019} this ripple in the beam width is shown to be asymmetric between the vertical and the horizontal direction (e.g. it is not rotation invariant). Our model can be considered a spherical averaged version of this effect which we believe still captures the main issues of such ripple. The combination of beam asymmetries and sky rotation will probably result in a superposition of sine waves which will leak the ripple across more scales while reducing its overall amplitude.

So far we have discussed the FWHM of the beam that we are assuming fully describes its frequency behavior. We address now the possible choices to describe how the beam behaves as a function of the polar angle $\theta$.
One interesting option for a beam model is the jinc function ${\rm jinc}(x) \equiv J_1(x)/x$,  where $J_1(x)$ is the Bessel function of the first kind. This model is quite popular as it corresponds to an unblocked circular aperture with uniform illumination \citep{wilson2013}. We use:
\begin{equation}
B_{\rm J}(\nu,\theta)=4\ {\rm jinc}^2\Big(\pi \frac{\theta}{\Delta\theta} \Big).
\label{eq:Jinc_beam}
\end{equation}
Note that the correct derivation would use $\sin{\theta}$ instead of $\theta$ in the equation above. They give similar results for $\theta < 10$ deg but will start to deviate after that. We found that using $\theta$ instead provides a better behavior at large angles.
We can see from figure \ref{fig:beam_models} that this function follows the beam main lobe accurately and captures the nulls of the first sidelobes. However, the amplitude of the sidelobes is higher than the MeerKAT/Eidos beam.

Another option with smaller sidelobes is the beam pattern resulting from a cosine-tapered field (or cosine-squared power) illumination function (\citealt{condonransom2016}),
\begin{equation}
B_{\rm C}(\nu,\theta)=\left[\frac{\cos{(1.189\theta\pi/{\Delta\theta})}}{1-4(1.189\theta/{\Delta\theta})^2}\right]^2.
\label{eq:Cos_beam}
\end{equation}

Figure~\ref{fig:beam_models} shows that this model fits the main lobe and the first two sidelobes quite well.
Indeed, a simplified MeerKAT beam model based on this function is publicly available\footnote{\href{https://github.com/ska-sa/katbeam}{https://github.com/ska-sa/katbeam}}. More details for MeerKAT can be found in \citet{mauch2020}.

Finally, we also consider a Gaussian function:
\begin{equation}
    B_{\rm G}(\nu,\theta)=e^{-4\ln(2)\left(\frac{\theta}{\Delta\theta}\right)^2}.
    \label{eq:Beam_gauss}
\end{equation}
This is the simplest case and we include it here for consistency. It is a good approximation to the main lobe (at least down to an order of magnitude) but it neglects completely the effects of the sidelobes. One important point is that all these beams are a function of the ratio $\frac{\theta}{\Delta\theta}$ so that a change in $\Delta\theta$ is equivalent to a rescaling in $\theta$.

A somewhat more accurate option was proposed in \citet{harper2018} based on the transformation of a Gaussian tapered airy disk. This model is also presented in figure~\ref{fig:beam_models} after tuning the parameters. 
We can see from the figure that within 0 to 1 degree (i.e. within the main lobe) all beams do match. Beyond that, the {\it Jinc} beam sidelobes drop slower in amplitude compared to the sidelobes of the other beams.
The airy beam has a trend with $\theta$ similar to the cosine but does not scale easily with frequency and the integration parameters need to be adjusted at each frequency. For these reasons, we focus the rest of our analysis on the {\it Cosine} beam as the best description of the MeerKAT true beam and retain the {\it Gaussian} and the {\it Jinc} respectively as an alternative best and worst case scenario for sidelobes.

\section{Sky model}\label{sec:skymodel}
In this section we describe the various components of our sky model: the 21~cm signal (section~\ref{sec:HI}), the Galactic and extra-galactic Free-free emission (section~\ref{sec:FF}), the Galactic synchrotron emission (section~\ref{sec:Gsynch}) and a new simulated Point Source catalogue (section~\ref{sec:ps}). This latter is a central ingredient to the analysis presented in this work.
Indeed, point source emission can have a non trivial impact in the cleaning when the standard assumption of a Gaussian beam is relaxed.
The other components are obtained using the publicly available code: Cosmological Realizations for Intensity Mapping Experiments (CRIME\footnote{\href{http://intensitymapping.physics.ox.ac.uk/CRIME.html}{http://intensitymapping.physics.ox.ac.uk/CRIME.html}}), presented in \citet{alonso2014}, where the interested reader can find more details. In the following we just give a brief overview. 
All the components of the sky model are given in full-sky HEALPix maps \citep{gorski2005} at selected frequencies and then combined and masked to mimic an IM survey with MeerKAT. 
We will discuss the simulated survey details later in section~\ref{sec:sky_area}.

\begin{figure}
\includegraphics[width=8.5cm]{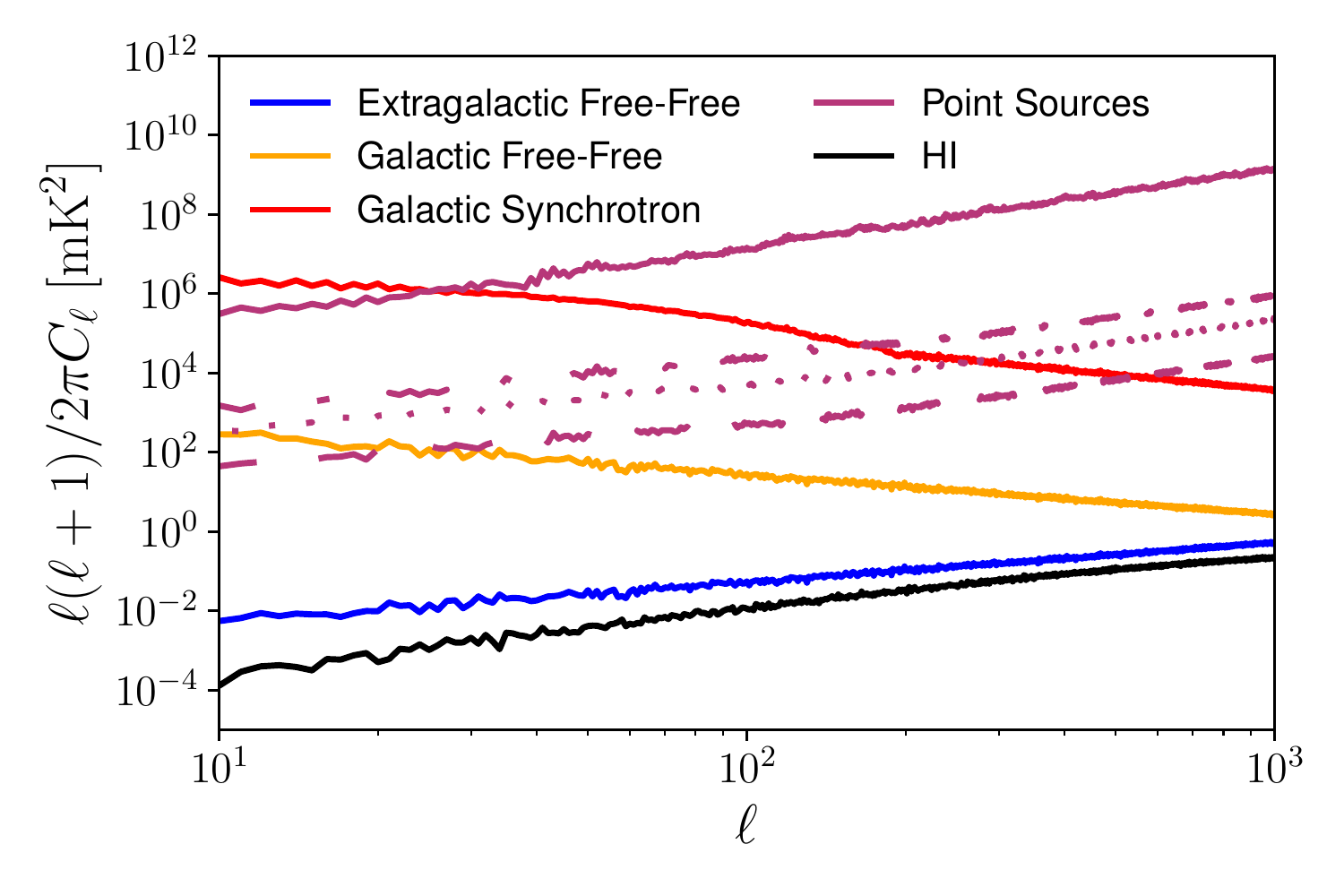}
\caption{Angular power spectrum of the HI signal and the the foregrounds at $950~\mathsf{MHz}$. The point source $C_\ell$ are shown for different flux cuts:  full catalogue (solid line), 10 $\mathsf{Jy}$ cut (dashed-dotted line), 1 $\mathsf{Jy}$ cut (dotted line) and 100 $\mathsf{mJy}$ cut (dashed line).}
 \label{fig:cl_fg}
 \end{figure}

\subsection{The 21~cm signal}\label{sec:HI}
To model the 21~cm signal we need to simulate the brightness temperature of the neutral hydrogen emission in every position on the sky and at every frequency of interest. There is a one to one relation between the observed frequency, $\nu$ , and the redshift $z$ of the emission: $\nu=\nu_{21}/(1+z)$, where $\nu_{21}$ is the frequency of the 21~cm line in its rest frame, $\approx 1420~\mathsf{MHz}$. 
The value of the 21~cm brightness temperature can be related to the underlying neutral hydrogen density  $\rho_{\rm HI}(\hat{n},z)$ through \citep{furlanetto2006}
\begin{equation}\label{eq:T21}
    T_{21{\rm cm}}(\hat{n},z)=0.19 \frac{\Omega_b h(1+z)^2}{\sqrt{\Omega_m(1+z)^3+\Omega_{\Lambda}}} x_{\rm HI}(z) \frac{\rho_{\rm HI}(\hat{n},z)}{\bar{\rho}_{\rm HI}(z)}\:{\rm K},
\end{equation}
where $\Omega_b$, $\Omega_m$, $\Omega_{\Lambda}$ are the baryon, total matter and dark energy density fractions respectively,
$x_{\rm HI}(z)$ is the neutral hydrogen mass fraction (with respect to baryons) and $\bar{\rho}_{\rm HI}(z)$ is the mean HI density at redshift z.

As already mentioned, the 21~cm signal is simulated using the CRIME code.
The code assumes the \citet{planck2013} best fit cosmology
and starts from a log-normal dark matter simulation on a cubic grid. 
The dark matter box is then divided in spherical shells, which are in
turn pixelized to yield 21~cm maps. A temperature is associated to each pixel considering the hydrogen density enclosed within it through equation~\ref{eq:T21}. Redshift distortions are also introduced using the velocity field.
The evolution of the neutral hydrogen fraction is assumed to be $x_{\rm HI}(z)=0.008 (1 + z)$, in agreement with the trend present in existing data \citep{crighton2015}. 
An example of the angular clustering of the signal is presented in figure~\ref{fig:cl_fg}.

\subsection{Free-free emission}\label{sec:FF}
The thermal bremsstrahlung emission produced by free electrons scattering off ions (free-free) originates from ionized hydrogen clouds (HII regions). Free-free can have both Galactic or extragalactic origin. 
In the simulations we use here, from \citet{alonso2014}, statistical isotropy is assumed. Such premise can break down, especially for the Galactic emission, which can be problematic if the fluctuations across the sky are strong, due to the convolution with the primary beam. However, we expect this signal to be a few orders of magnitude below the galactic synchrotron and therefore have a subdominant effect.
Full-sky free-free maps are therefore generated, following \citet{santos2005}, as a Gaussian realization of
\begin{equation}\label{eq:SCK}
 C_{\rm \ell} (\nu_{1}, \nu_{2}) = A \left(  \frac{\ell_{\rm ref}}{\ell}  \right)^{\beta}  \left( \frac{ \nu_{\rm ref}^{2}}{ \nu_{1}\nu_{2}} \right)^{\alpha}  {\rm exp} 
 \left( - \frac{{\rm log^{2}} ( \nu_{1} / \nu_{2} ) }{2 \xi^{2}} \right),
\end{equation}
where $A$ denotes the overall amplitude and $\beta$ parameterize the foreground distribution on angular scales.
The parameter $\alpha$ is the foreground spectral index and $\xi$
is the frequency-space correlation length that parameterizes the characteristic frequency scale over which foregrounds are correlated.
The parameter values can be found in table~\ref{tab:alp}. 
As before, we make use of the code presented in \citet{alonso2014} to create the desired foreground maps. 

\begin{table}
\centering
  \caption{Foreground $C_{\ell}(\nu_1, \nu_2)$ model parameters for the pivot values $\ell_{\rm ref}=1000$ and
  $\nu_{\rm ref}=130 \ \mathsf{MHz}$ \citep{santos2005}}
  \begin{tabular}{c c c c c}
  \hline
 Foreground & $A \ \mathsf{[mK^2]}$ & $\beta$ &  $\alpha$ & $\xi$ \\
  \hline
  Galactic free-free & 0.088 & 3 & 2.15 & 35 \\
  Extragalactic free-free & 0.014 & 1 & 2.1 & 35 \\
  Galactic synchrotron  & 700 & 2.4 & 2.8 & 4 \\
  \hline
  \label{tab:alp}
\end{tabular}
\end{table}

\subsection{Galactic Synchrotron}\label{sec:Gsynch}
Galactic synchrotron radiation is the strongest foreground emission in the frequency range of interest for intensity mapping and is produced by cosmic-ray electrons and positrons propagating in interstellar magnetic fields. 
Since a full physical model would require the knowledge of magnetic fields, cosmic-ray  electron  distributions and  propagation, the standard approach is to resort to data driven modelling that relies on the Haslam full-sky map at $408~\mathsf{MHz}$ \citep{haslam1982}.
In \citet{alonso2014}, the Haslam map is extrapolated to the frequency range of interest 
using a spectral index $\alpha(\hat{n})$ taken from the Planck Sky Model \citep[PSM,][]{delabrouille2013}. 
Moreover, since the Haslam map has poor resolution, small angular scales ($\ell>200$) are filtered out and replaced with the isotropic model of \citet{santos2005} discussed in the previous section and whose parameters are reported in table \ref{tab:alp}. An example of angular clustering of the galactic synchrotron emission at fixed frequency is reported in figure~\ref{fig:cl_fg}, where the transition between the two regimes can be noticed.

\begin{figure}
 \centering
 \includegraphics[scale=0.42]{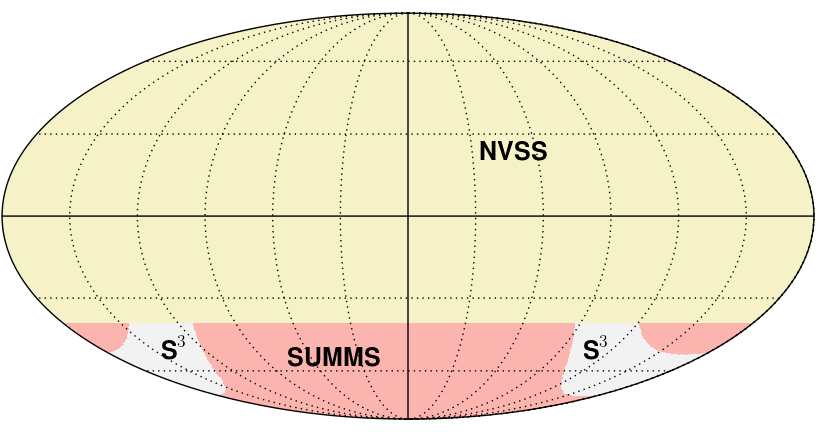}
 \caption{Survey coverage for NVSS and SUMSS. The yellow area indicates the NVSS coverage while the SUMMS area is in pink. The grey area is the SUMSS masked area $i.e.$ galactic latitude $|b|<$10$^{\circ}$ which was filled with S$^{3}$ sources. Note that below a flux cut of 5 $\mathsf{mJy}$ (see text for details) the full sky is filled with S$^{3}$ sources.}
 \label{fig:ht1}
\end{figure} 

\subsection{Point sources}\label{sec:ps}
We now focus on the extragalactic point source (PS) modelling.
Previous studies \citep[e.g.][]{santos2005,alonso2014} have considered PS as a statistically isotropic emission that can be described with a frequency-space angular power spectra.
In this work, we propose a more realistic model for the point sources that exploits available data and well validated simulations to create a full-sky catalogue at $1.4$ GHz, from which we produce pixelized maps in the frequency range of interest. Such PS model will be particularly relevant for the primary beam effects we are studying.

\subsubsection{Full sky PS catalogue}\label{sec:ps_cat_I}
Our starting point is the SKA Simulated Skies (S$^{3}$, \citealt{wilman2008}), a semi-empirical simulation of the extragalactic radio continuum whose simulated sources, that include AGN and star-forming galaxies, are drawn from realistic luminosity functions and follow with appropriate bias the underlying dark matter density field. 
The minimum flux density, $S_{\rm min}$ obtained from S$^{3}$ is 10 $\mu \mathsf{Jy}$.
The S$^{3}$ simulation covers a patch of 20$\times$20 deg$^{2}$. To extend the catalogue to full sky we use the following approach:
\begin{itemize}
\item we consider the S$^{3}$ area, $\Omega_{\rm sim}$ and group all the flux densities into logarithmic bins. Every flux bin $i$ contains $N_{\rm sim}^{i}$ sources;
\item we create an empty HEALPix map with $N_{\rm side}=512$, whose pixel area is $\Omega_{\rm pix}$\footnote{Note that $N_{\rm side}=512$ corresponds to $\Omega_{\rm pix}\sim 0.01\,{\rm deg}^2$.}. The resolution of this map will give the resolution of our final catalogue;
\item we assume that the number of sources scales linearly with area such that the mean number of sources per pixel corresponding to the i-th flux bin is $\bar{N}_{\rm pix}^{i}$ = $N_{\rm sim}^{i} \left(\frac{\Omega_{\rm pix}}{\Omega_{\rm sim}}\right)$;
\item we assign to every pixel a certain number of sources for every flux bin, Poisson sampled from $\bar{N}_{\rm pix}^{i}$;
\item we assign to every source a spectral index, $\alpha$, drawing from a Normal distribution $\mathcal{N} (-0.8, 0.2)$, in broad agreement with \citet{Garn2008}. This spectral index dictates the frequency scaling of the source flux, using the standard power law $S \propto \ \nu^{\alpha}$;
\item right ascension and declination of every source are obtained from the pixel position in the sphere. Note that multiple sources will have the same (ra,dec).  This is not a limitation as long as the map resolution is higher than our target experimental resolution.
\end{itemize}
With this procedure we construct a full-sky catalogue of S$^3$-like sources. We note that this catalogue does not include angular clustering seen in real data (e.g. \citealt{de_Oliveira_costa_apr2010, de_Oliveira_costa_jun2010}).
To make the catalogue more realistic, we therefore added the observational data from two radio surveys, that naturally carry the information on the large-scale distribution of matter. 
We consider the data from the National Radio Astronomy Observatory VLA Sky Survey (NVSS, \citet{condon1998}) at frequency $1.4$~GHz and the Sydney University Molonglo Sky Survey (SUMSS, \citet{mauch2003}) with frequency $843\mathsf{MHz}$. NVSS covers declination $\rho \geqslant \ -40^{\circ}$ in  while SUMSS covers $\rho \leqslant \ -30^{\circ}$.  
To avoid double counting of sources we only consider SUMSS at declination $\leqslant \ -40^{\circ}$. Note also that SUMSS has a galactic mask at latitude $|b|< 10^{\circ}$. Inside this area we keep using the S$^3$ generated sources.

\begin{figure}
    \centering
    \includegraphics[width=8cm]{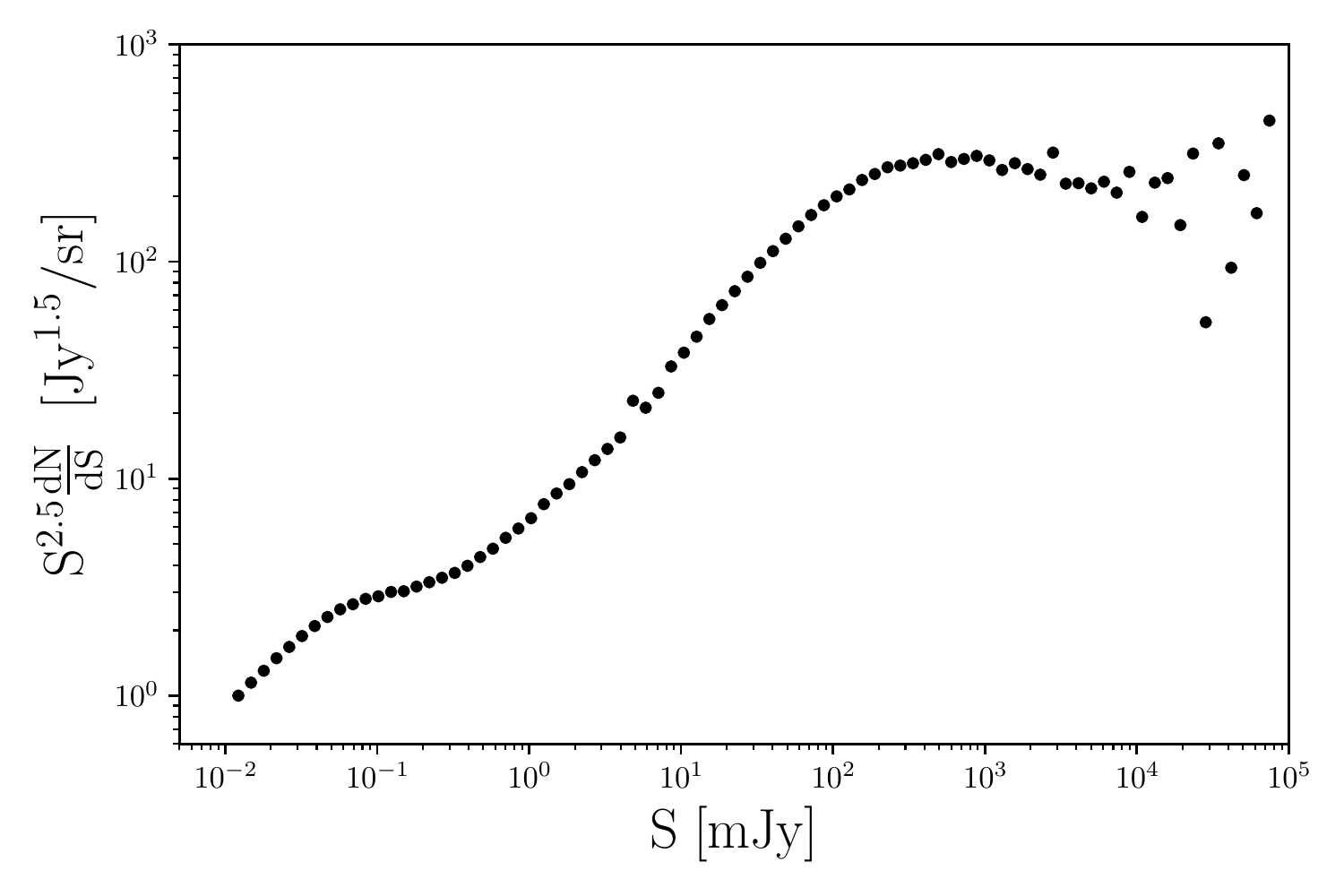}
    \caption{The normalized intrinsic source counts plot of the catalogue at 1.4 $\mathsf{GHz}$. Below 5 $\mathsf{mJy}$ are the sources from the S$^{3}$ simulation while above are NVSS sources, extrapolated SUMSS sources or S$^{3}$ (see figure~\ref{fig:ht1}).}
    \label{fig:dfs}
\end{figure}

The resulting sky coverage of the NVSS, SUMMS and S$^3$ is shown in figure~\ref{fig:ht1} in equatorial coordinates.
Note that to extrapolate to other frequencies, we assign a spectral index to NVSS and SUMSS sources the same way as for the S$^3$ sources. We then use these spectral indexes to extrapolate SUMSS flux densities from $843~\mathsf{MHz}$ to the common $1.4~~\mathsf{GHz}$ frequency.
For both NVSS and SUMSS we only keep the sources higher than a specific flux cut, $S_{\rm cut}$, at $1.4$~GHz to avoid spurious detections.
We set $S_{\rm cut}=10\sigma_{\rm NVSS}\ \simeq \  5\ \mathrm{mJy}$ where $\sigma_{\rm NVSS}$ is the NVSS noise flux rms.
On the NVSS and SUMMS sky patches, below $S_{\rm cut}$, we include the S$^3$ sources. One may worry that below $S_{\rm cut}$ the point sources are not correlated across the sky. However, for the objectives of this paper, this lack of correlation should not affect the conclusions from the beam effects.
In figure~\ref{fig:dfs} we reconstruct the differential source counts diagram at 1.4~GHz. 
The scatter at high flux density is due to Poisson sampling with a low average number of sources per bin.
The catalogue can be downloaded from the \href{https://drive.google.com/drive/folders/1bbm7ExfkA1-jT-o8yme5DaeXMWOrjBTd?usp=sharing}{CRC repository}. A detailed description of its structure can be found in appendix~\ref{app:ps_cat_P}, together with the description of the polarized part of the catalogue which is not being used in this analysis.

\subsubsection{Point source maps}\label{sec:ps_maps}
\begin{figure}
 \centering
 \includegraphics[scale=0.42]{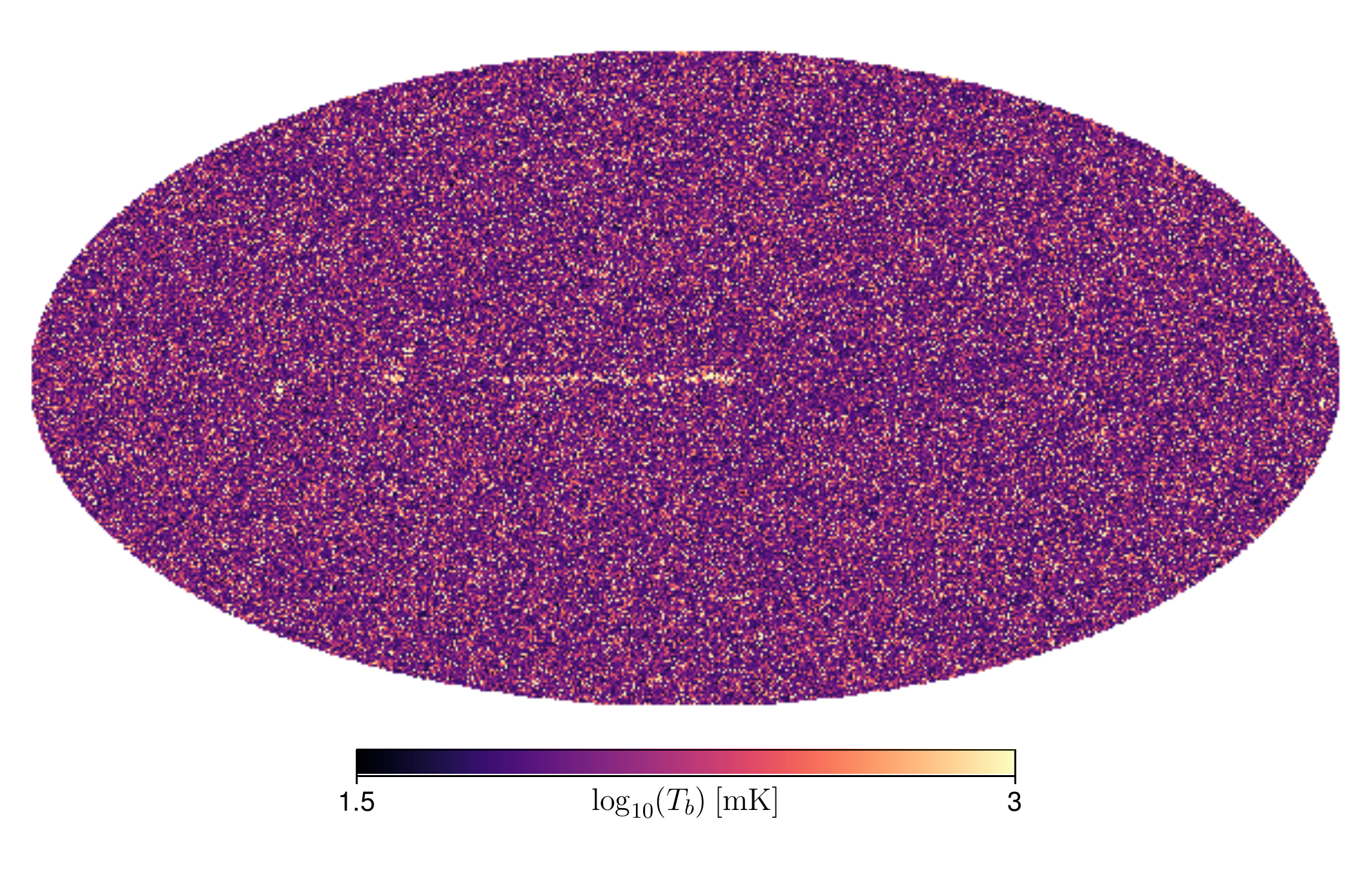}
 \caption{Our PS model at 950~$\mathsf{MHz}$ in galactic coordinates.}
 \label{fig:ps_map}
\end{figure}

For our simulation pipeline we need to transform the point source catalogue discussed in the previous section into a pixelized point source temperature brightness HEALPix maps at every frequency of interest. We use $N_{\rm side}=512$. As we will discuss further later on, this is a safe choice since the expected resolution of single-dish survey is a factor 10 worse. Given $N_{\rm side}$, the pixel area, $\Omega_{\rm pix}$, is fixed and the coordinates of every source can be associated to a specific pixel. The temperature brightness value for every pixel can then be computed using the Rayleigh-Jeans approximation:
\begin{equation}
      T^j_b(\nu)=\Bigg(\frac{c^2}{2 k_{\rm B} \nu^2 \Omega_{pix}}\Bigg)\sum_{i=1}^{N^j_p} S_i(\nu),
\end{equation}
where $k_{\rm B}$ is the Boltzmann constant, $c$ the speed of light and the sum is over all the sources falling in the j-th pixel. $S_i(\nu)$ is the flux density of each these sources rescaled from $1.4$~GHz to $\nu$ using the source spectral index.
An example of such map is given in figure~\ref{fig:ps_map}. 
In figure~\ref{fig:cl_fg} we show instead an example of the angular power spectrum of the same map compared to the other foreground emissions and the HI signal. Together with the $C_\ell$ of the map created using the full point source catalogue, we show how the angular power spectrum reduce in amplitude if successive stronger flux cuts are applied to the catalogue.

\section{Beam convolution}\label{sec:beamconv}
To simulate how a measured sky brightness temperature map will look like, we need to convolve the sky map with one of the beam models described in section~\ref{sec:beam}:

\begin{equation}\label{eq:Tmes}
    \tilde{T}_{\rm sky}(\nu,\theta,\phi)=\int 
    \left[ \hat{R}(\theta,\phi) B \right](\nu, \theta',\phi')T_{\rm sky}(\nu, \theta',\phi')d\Omega' , 
\end{equation}
where $d\Omega'=\sin{(\theta')}d\theta'd\phi'$ and $\hat{R}$ is the operator of finite rotations such that $\hat{R}B$ is the rotated beam into the $(\theta,\phi)$ direction \citep[e.g.][]{wandelt2001}.
The direct computation of this convolution integral can be avoided moving to spherical harmonic space.
A sky map can be decomposed into spherical harmonics,
 \begin{equation}\label{eq:Bt_sht}
 T(\nu,\theta,\phi) = \sum^{\ell_{max}}_{\ell=0}\sum^{\ell}_{m=-\ell}a_{\ell m}(\nu) Y_{\ell m}(\theta,\phi),
 \end{equation}
 \noindent where $Y_{\ell m}(\theta,\phi)$ are the spherical harmonic functions and $a_{\ell m}(\nu)$ are the spherical harmonic coefficients,
 \begin{equation}\label{eq:alm}
 a_{\ell m}(\nu) =\int T(\nu,\theta,\phi)Y^{\ast}_{\ell m}(\theta,\phi) d\Omega,
 \end{equation}
For a symmetrical beam this simplifies to 
  \begin{equation}\label{eq:bl}
 B(\nu,\theta)=\sum_{\ell}b_{\ell}(\nu)Y_{\ell 0}(\theta,\phi),
 \end{equation}
where the beam harmonic coefficients $b_l$ do not depend on $m$ and can be written as 
\begin{equation}\label{eq:b_l}
    b_{\ell}(\nu)=\int B(\nu,\theta,\phi)Y^{\ast}_{\ell 0}(\theta,\phi) d\Omega.
\end{equation}

The convolution theorem transforms the integral in a simple product in harmonic space. We therefore use it for including the effect of the beam in our simulated sky maps. At each frequency, we compute the spherical harmonic transform of the sky temperature maps and the one of the beam model projected onto a HEALPix map, using healpy routines \citep{zonca2019}. Then, a fast, element by element multiplication is performed, 
\begin{equation}\label{eq:conv_alms}
\tilde{a}_{\ell m}(\nu) =\sqrt{\frac{4\pi}{2\ell+1}}a_{\ell m}(\nu)\frac{b_{\ell}(\nu)}{\sqrt{4\pi}b_0(\nu)}.
\end{equation}
Note that we are assuming that the beam function, $B(\nu,\theta,\phi)$, is defined to be one at the center. We then need to divide by $\sqrt{4\pi}b_0(\nu)$ so that the beam integrated over the sky is normalized to 1. This is the required normalization in order to recover the signal angular power spectrum with the correct amplitudes.
To obtain the beam convolved sky maps we simply use $\tilde{a}_{\ell m}(\nu)$ in equation~\ref{eq:Bt_sht}.

\subsection{Beam effects on a single point source}\label{sec:single_ps}

To examine the effect of a frequency dependent beam, we start by looking at the simplified case of a single point source.
Since this point source should be represented by a Dirac delta function in terms of the sky temperature, the measured brightness temperature is:
\begin{equation}
      T_P(\nu,\theta)=\frac{\lambda^2}{2k_{\rm B} \int B(\nu,\Omega) d\Omega}S(\nu)B(\nu,\theta),
\end{equation}
where $\theta$ is the angle of the point source with respect to the beam pointing and, again, $B(\nu,\theta)$ is the telescope beam normalized to 1 at the center. As discussed in section~\ref{sec:beam}, our standard description for the MeerKAT beam is the Cosine model of equation~\ref{eq:Cos_beam} and we use the full frequency dependence in equation~\ref{eq:fwhm_nu}. For simplicity, we take the point source flux, $S(\nu)$, to be constant in frequency and equal to 1 Jy. This corresponds to a temperature contribution of about 0.05 K at the peak.

\begin{figure}
\includegraphics[width=9cm]{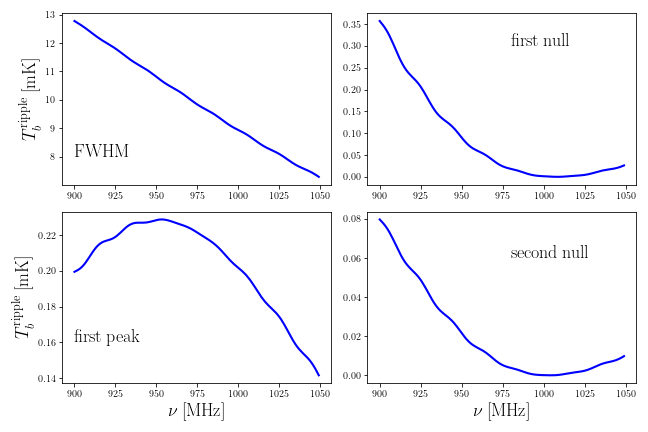}
\includegraphics[width=9cm]{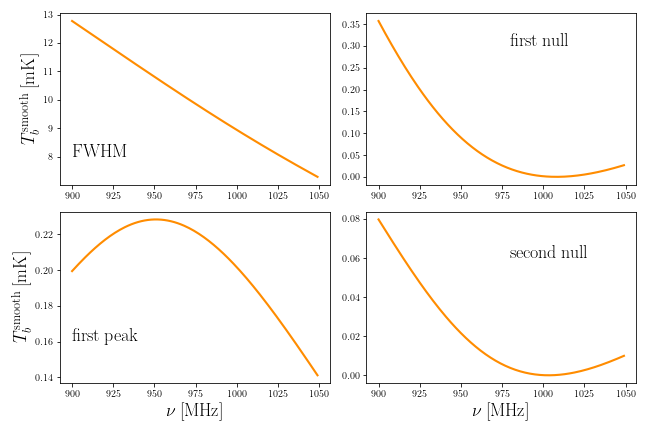}
\includegraphics[width=9cm]{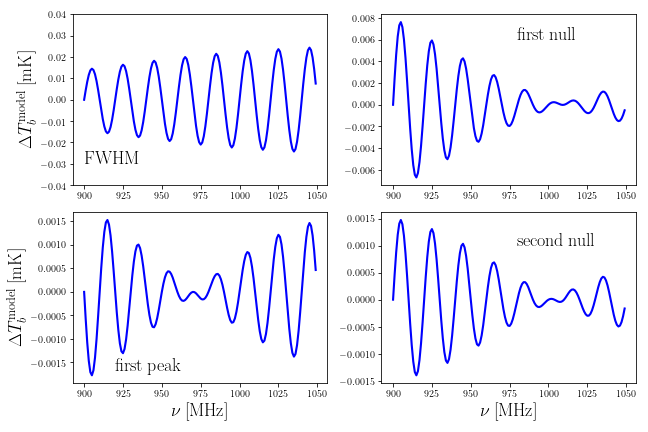}
\caption{{\it Top panel:} Behavior of $T_b(\nu)$ for a single point source in different angular positions with respect to the center of the beam as a function of frequency. The positions are defined at 1~GHz. The beam model considered is the Cosine beam with $\Delta \theta$ of equation~\ref{eq:fwhm_nu} (ripple model). {\it Central panel:} same as before but considering the the Cosine beam with only the polynomial part of $\Delta \theta$ of equation~\ref{eq:fwhm_nu} (i.e. with $A=0$, smooth model).
{\it Bottom panel:} The residual behavior of $T_b(\nu)$ for the ripple model after the smooth component is subtracted.}
 \label{fig:ripple_effect_single_PS}
 \end{figure}

The upper panel of Figure~\ref{fig:ripple_effect_single_PS} shows the effect of the convolution with a frequency dependent beam ({\it ripple} model) for a few positions of the point source with respect to the beam center. The angle positions are defined for the standard "$\lambda/D$" case at 1~GHz (first peak refers to the first sidelobe peak). The {\it smooth} component of the beam model (central panel) creates a the slowly varying behavior near the peaks and nulls of the sidelobes but we expect the foreground cleaning algorithms to be able to deal with this type of structure. Of course, the situation can become more complicated when we combine several strong point sources with different spectral indexes. Still, the overall effect is reasonably benign as we will see later. 

Assuming that we know the beam well enough to remove the effect from the smooth component, we are left with the residuals seen in the bottom panel of figure~\ref{fig:ripple_effect_single_PS} which are much harder to deal with. If no further cleaning can be done, this will give the final contribution from this single point source. Note that we cannot remove this effect through a gain calibration as the amplitude of the effect changes with $\theta$ (and is zero at the peak). Interestingly, although the relative effect is stronger near the nulls, a point source at the FHWM gives a stronger overall contribution with a similar shape. However, the signal is around 0.02 mK. Since the 21cm signal rms is about $0.1$ mK, this means that a 1 Jy source would have an impact at the 20\% level at most. Again, the situation will complicate once we include more point sources. The strength of the signal goes down as we move away from the center but there is a higher chance of finding more and stronger point sources. A 10 Jy source on the second null would generate contamination at the 10\% level. Although we have been considering a single point source, the same analysis could be applied to fluctuations across the sky from diffuse components (a completely smooth component would not suffer from beam effects). A 0.1 K fluctuation over degree scales would show up in the first null and also contribute at the 10\% level. Such fluctuations are expected with the galactic synchrotron. 

This was just a basic calculation to show the expected contamination level, but a proper simulation is needed to take all the effects into account as we will in section~\ref{sec:results_freq}. On the one hand, the combination of more sources across the sky will make the cleaning more complex. On the other hand, the cleaning algorithm might be more robust. Moreover, techniques such as point source subtraction and beam deconvolution can improve the overall outcome.

\section{Simulation pipeline}\label{sec:pipe}
In this section we describe our simulated data, detailing the survey specification in section~\ref{sec:sky_area}, tailored to MeerKLASS \citep{santos2017}, a proposed wide area survey with the
MeerKAT telescope. We then detail the contents of the final maps in section~\ref{sec:flow} and present the foreground subtraction strategy in section~\ref{sec:blind}. The estimators used to quantify the signal recovery are presented in section~\ref{sec:fg_diag}.

\subsection{Survey specifications}\label{sec:sky_area}
\begin{figure}
 \centering
 \includegraphics[scale=0.4]{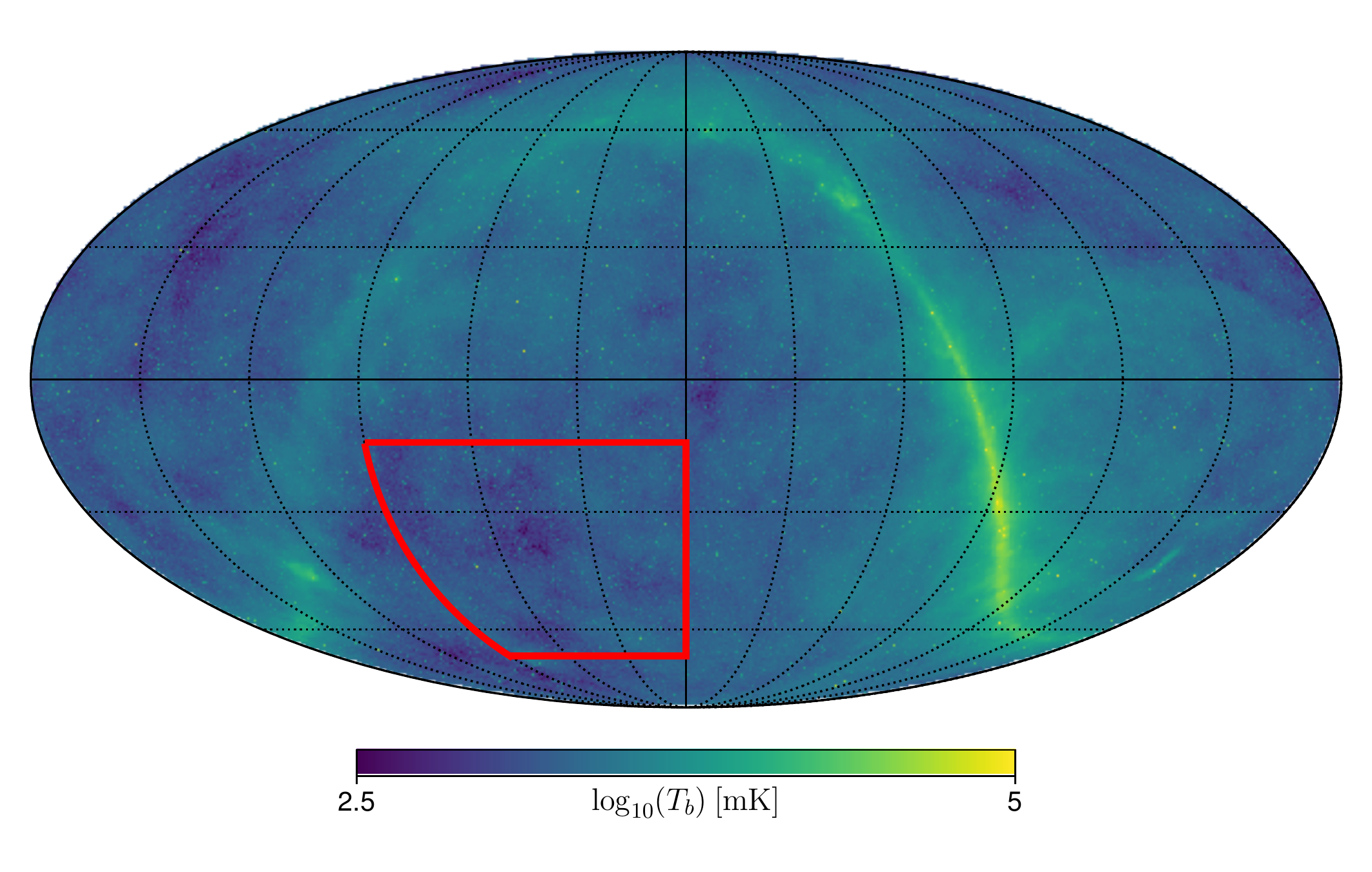}
 \caption{The simulated survey region in equatorial coordinates (red contours) covering $~4k$~deg$^2$ of the sky ($f_{\rm sky}\sim 0.09$). For illustrative purpose we show in the background the simulated foreground emission at $950~\mathsf{MHz}$. The simulated survey area avoids the strong emission coming from the Galactic plane.}
 \label{fig:mask}
\end{figure}

Following what was proposed with the MeerKLASS survey, we consider a survey area of almost 10\% of the sky,
overlapping with the sky area probed by the Dark Energy Survey (DES), in order to allow comparative studies and cross-correlation analysis that will ultimately help in constraining cosmological parameters.
We choose the sky patch as in figure~\ref{fig:mask} and consider observations between $900~\mathsf{MHz}$ and $1050~\mathsf{MHz}$ with a frequency resolution of 1~$\mathsf{MHz}$. The 150 channels cover the redshifts range $z \ \in$ (0.35, 0.58). We store the mock data in HEALPix at $N_{\rm side}=512$ that corresponds to a pixel resolution of $\theta_{pix}=0.11$~deg. 

Table~\ref{Tab:survey_spec} summarizes the survey specifications and instrumental parameters.
The instrumental noise can be computed as function of these parameter choices.
In our simulations we consider only thermal noise, that is, Gaussian noise with null mean and a standard deviation $\sigma_{N}$ \citep{wilson2013}:
\begin{equation}\label{eq:RMS}
 \sigma_{N} = \frac{T_{\rm{sys}}}{\sqrt{2 t_{\rm pix} \Delta \nu  }},
\end{equation}
with $T_{\rm sys}$ the system temperature, $\Delta{\nu}$ the frequency resolution and
$t_{\rm pix}$ the total integration time spent on a single pixel,
\begin{equation}
t_{\rm pix} =  t_{\rm obs}  N_{\rm dish} \frac{ \Omega_{\rm pix} } { \Omega_{\rm sur} }.
\end{equation}
Where $t_{\rm{obs}}$ is the total integration time, $\Omega_{\rm{sur}}=4 \pi f_{\rm sky}$ the survey area, $ \Omega_{\rm{pix}}$ the pixel area and $N_{\rm{dish}}$ the number of telescope dishes\footnote{Note that, although data are auto-correlation of single dish measurements, the final maps will be a combination of the independent measurements obtained with the various dishes, enhancing the signal-to-noise.}.
With the assumptions listed in table~\ref{Tab:survey_spec} and the choice of  $N_{\rm side}=512$
we obtain a noise rms value of 0.245 $\mathsf{mK}$.
The resulting noise map $T^{f_{sky}}_{{\rm noise}}( \hat{\mathbf{n}})$ takes into account the sky patch considered and every unmasked pixel contains values drawn from a Gaussian distribution. Note that in principle $T_{{\rm sys}}$ should include a frequency dependent evolution inherited from the sky temperature, but we are neglecting this small variation and assuming the noise to be constant with frequency. We do not expect the specifics of the instrumental parameters assumed here to affect the main conclusions of the paper, which can be easily extrapolated to the SKA.

\setlength{\arrayrulewidth}{0.1mm}
\begin{table}
   \centering
   \caption{Instrumental parameters used to simulate an intensity mapping survey with MeerKAT.}
    \begin{tabular}{c  c}
    \hline
   $D$ (dish diameter)  &  13.5 m \\
   $t_{\rm obs}$  & 4 000 h \\
   $\Delta \nu$  & 1 $\mathsf{MHz}$ \\
   $N_{\nu}$    & 150 \\
   $N_{\rm dish}$ & 64 \\
   $T_{\rm sys}$  & 20 $\mathsf{K}$ \\
  ($\nu_{\rm min}$, $\nu_{\rm max}$) & (900, 1050) $\mathsf{MHz}$ \\
  ($z_{\rm min}$, $z_{\rm max}$) & (0.35, 0.58) \\
  $f_{\rm sky}$ & 0.09 \\
  survey area ($\Omega_{\rm sur}$) &  $\sim$ 3 700 deg$^{2}$\\
  \hline
 \label{Tab:survey_spec}

 \end{tabular}
\end{table}

\subsection{Mock final IM maps}\label{sec:flow}
To construct our mock sky maps, we start by adding together all the elements described in section~\ref{sec:skymodel}, at every frequency $\nu$,
\begin{equation}\label{eq:total_intensity}
T_{{\rm sky}}(\nu,\hat{\mathbf{n}})=T_{{\rm HI}}(\nu,\hat{\mathbf{n}})+T_{{\rm fg}}(\nu,\hat{\mathbf{n}})
\end{equation}
where $T_{{\rm HI}}(\nu,\hat{\mathbf{n}})$ is the brightness temperature of the 21~cm signal, constructed as described in section~\ref{sec:HI} using a 1~Gpc$^3$ box with $N_{\rm grid}=2048$ and
\begin{equation}\label{eq:fg}
T_{{\rm fg}}(\nu,\hat{\mathbf{n}})=T_{{\rm PS}}(\nu,\hat{\mathbf{n}})+T_{{\rm Gsynch}}(\nu,\hat{\mathbf{n}})+T_{{\rm GFF}}(\nu,\hat{\mathbf{n}})+T_{{\rm EGFF}}(\nu,\hat{\mathbf{n}}),
\end{equation}
where $T_{{\rm PS}}$ represents the brightness temperature of the point sources, $T_{{\rm Gsynch}}$ is the Galactic synchrotron emission and $T_{{\rm GFF}}$, $T_{{\rm EGFF}}$ are the Galactic and extra-Galactic free-free emission.

The sky signal needs to be convolved with the beam and masked 
to mock the MeerKAT observations. The beam convolution is performed as described in section~\ref{sec:beamconv}, i.e. at the full sky level to avoid complication at the edges of the mask. Our reference mock maps include the Cosine beam model described in section~\ref{sec:beam}, but we produce maps using also the Gaussian and the Jinc beam for comparison. For every beam model we need to specify its frequency dependency: either the simple proportionality to $\lambda/D$ of equation~\ref{eq:fwhm}, or the {\it smooth}/{\it ripple} model described in equation~\ref{eq:fwhm_nu}.
For each of these beam models we compute the spherical harmonic coefficients $b_\ell$ defined in equation~\ref{eq:bl} and produce a final convolved map $\tilde{T}_{{\rm sky}}(\nu,\hat{\mathbf{n}})$ 
using equation~\ref{eq:Bt_sht} and \ref{eq:conv_alms}.
The final full sky map is then masked to the target area and the thermal noise is added. 
\begin{equation}\label{eq:final_conv_map}
     T_{{\rm obs}}^{f_{ sky}}(\nu, \hat{\mathbf{n}})=\tilde{T}^{f_{sky}}_{sky}(\nu,\hat{\mathbf{n}})+ T^{f_{sky}}_{{\rm noise}}(\hat{\mathbf{n}}).
\end{equation}
This final mock data product will be the input of the foreground cleaning algorithm.

\subsection{Foreground subtraction}\label{sec:blind}
The next step is the application of the foreground cleaning algorithm. Showing how the beam effects will affect such cleaning is one of the main goals of the paper. 
The foregrounds are expected to be spectrally smooth (e.g. highly correlated in frequency), contrary to the 21~cm signal (and the noise). The underlying assumption to all blind cleaning methods is however just that the foreground emission is very large in amplitude.
In this work, following an already successful applications path in IM studies \citep{wolz2014,alonso2015,wolz2017,cunnington2019,carucci2020}, we apply {\it blind} methods for foreground subtraction.
We use both the Principal Componet Analysis (PCA) and Fast Independent Component Analysis (FastICA) algorithm \citep{fastica} to solve for the foreground components, using the code presented in \citep{alonso2015}.
For all our cases of study PCA and FastICA yield almost identical results. For this reason, we present our results only for the PCA algorithm.

The PCA cleaning method relies on taking the strongest $N_{\rm fg}$ eigenvalues of the frequency-frequency covariance matrix of the data, which should be coming from the foreground components. 
This number of foreground components, $N_{\rm fg}$, depends on the complexity of the problem and is a free parameter in the code. In our context, for example, a first guess could be $N_{\rm fg}=4$ as four are the different foregrounds (galactic synchrotron, point sources and galactic and extra-galactic free-free), but it is possible that a smaller/higher amount of components would be needed. If $N_{\rm fg}$ is too small we will under-clean and confuse the signal with residuals from foregrounds and instrumental effects. In contrast, if $N_{\rm fg}$ is too large, we will start removing the actual 21cm signal,  irreparably compromising its interpretation. Moreover, it is expected that the convolution with the beam will play a role in defining the optimal number. Establishing the number of components to remove thus requires a careful study with simulations.

\subsection{Foreground cleaning diagnostics}\label{sec:fg_diag}
To assess the degree of success of the foreground cleaning procedure we will use two estimators: the angular power spectrum, which describes the clustering on the angular direction at fixed frequency, and the radial power spectrum which instead describes the clustering along the line of sight.

\subsubsection{Angular power spectrum}
We define the temperature brightness fluctuation contrast $\Delta T$ as the difference between the temperature in each pixel and the mean of the sky patch under consideration.
At every frequency, the angular power spectrum in the full sky case can be estimated from the spherical harmonic coefficient $a_{\ell m}(\nu)$ of $\Delta T(\nu)$ using
\begin{equation}\label{eq:cl}
    \hat{C}_{\ell}(\nu) \equiv \frac{1}{2\ell+1}\sum_{m=-\ell}^{+\ell} |a_{\ell m}|^2 
\end{equation}
This estimator is no longer valid for sky patches, and finding the correct one is a non trivial problem. One widely used solution is to apply the 
Monte Carlo Apodized Spherical Transform Estimator \citep[MASTER,][]{hivon2002}
to correct the $C_{\ell}$ for the effect of the mask.
When the telescope scanning strategy probes a small patch of the sky with sharp edges, recovering the signal is also highly non trivial for the MASTER estimator.

 Given that for our purposes we are interested in the quality of the cleaning and not in the shape of the signal itself, we will compute the $C_{\ell}$ using simply equation~\ref{eq:cl}, correcting only for the sky fraction (dividing by $f_{\rm sky}$). The resulting $C_{\ell}$ will have approximately the same trend and amplitude as the full-sky ones but will display the well known oscillatory behavior due to the coupling of different scales induced by the presence of the mask. The HI signal with which we compare will suffer the same issue since the signal angular power spectrum is also computed on the selected sky patch, justifying the direct use of equation~\ref{eq:cl}.

To show the quality of the PCA cleaning we will also use the quantity:
\begin{equation}\label{eq:Dcl}
  \langle (C_{\ell}^{\rm rec}-C_{\ell}^{\rm true})/C_{\ell}^{\rm true}\rangle \equiv \langle \Delta C_{\ell}/C_{\ell}^{\rm true} \rangle,  
\end{equation}
where $\langle.\rangle$ indicates averaging over the frequency channels.

\subsubsection{Radial power spectrum}
To explore clustering along the line of sight we use the definition of the radial power spectrum as in \citet{alonso2014}. The frequency band is divided into slices, within which the universe should not evolve significantly and a constant redshift, $z_{\rm eff}$, is assumed. The slices, however, should be big enough for capturing all the relevant scales. We then compute the Fourier transform of the temperature fluctuations per bin along each line of sight, $\tilde{\Delta T}(k_\parallel,\mathbf{\hat{n}})$, where
\begin{equation}
    k_\parallel=\frac{\nu_{21} H(z_{\rm eff})}{(1+z_{\rm eff})^2}k_{\nu}
\end{equation}
and $k_{\nu}$ is the Fourier conjugate of the frequency, e.g. $\delta k_{\nu}=2\pi/\Delta \nu$.
The radial power spectrum results from an average over all lines of sight in the sky patch,
\begin{equation}\label{eq:Pk}
  P(k_\parallel) = \frac{\Delta \chi}{2\pi N_{\rm pix}}\sum_{i=1}^{N_{\rm pix}}|\tilde{\Delta T}(k_\parallel,\mathbf{\hat{n}})|^2,
\end{equation}
where ${\Delta \chi }_s=\chi(z_s^{\rm max})- \chi(z_s^{\rm min})$ is the slice width.

Foreground removal methods generally struggle at the edges of the input frequency band. To partially avoid this bias we do not include in the analysis the $10~\mathsf{MHz}$ at the beginning and at the end of our frequency range. Given the small redshift range of our simulations, we use a single redshift bin corresponding to $z_{\rm eff}\sim 0.46$.
In section~\ref{sec:results} we will show the recovered $P(k_\parallel)$ for our cases of study. Moreover, as for the angular power spectrum we will consider the estimator
\begin{equation}\label{eq:Dpk}
(P(k_\parallel)-P^{\rm true}(k_\parallel))/P^{\rm true}(k_\parallel) \equiv \Delta P(k_\parallel)/P^{\rm true}(k_\parallel),
\end{equation}
to better assess the quality of the foreground cleaning.
We anticipate that, due to the smooth frequency dependence of the foregrounds, the cleaning is expected to inevitably compromise the largest radial scales.

\section{Results}\label{sec:results}
We discuss here the results of our simulation and cleaning pipeline. The mock observations finalized in section~\ref{sec:flow} are the input of the foreground cleaning method described in section~\ref{sec:blind}. To quantify the performance of the cleaning, we make use of the estimators detailed in section~\ref{sec:fg_diag}. The main focus of this work is to investigate how a non trivial shape and/or frequency dependence for the telescope beam impacts the cleaning. To this aim, we discuss the effect of sidelobes with particular attention to the role of the point source contamination in section~\ref{sec:results_sidelobes}, and analyze the increasing difficulties of the cleaning procedure in the presence of a frequency dependent FWHM in section~\ref{sec:results_freq}.
In section~\ref{sec:resmooth} we investigate how smoothing the maps to a common resolution could help the cleaning process.

\subsection{The effect of beam sidelobes}\label{sec:results_sidelobes}

\begin{figure}
\includegraphics[width=8cm]{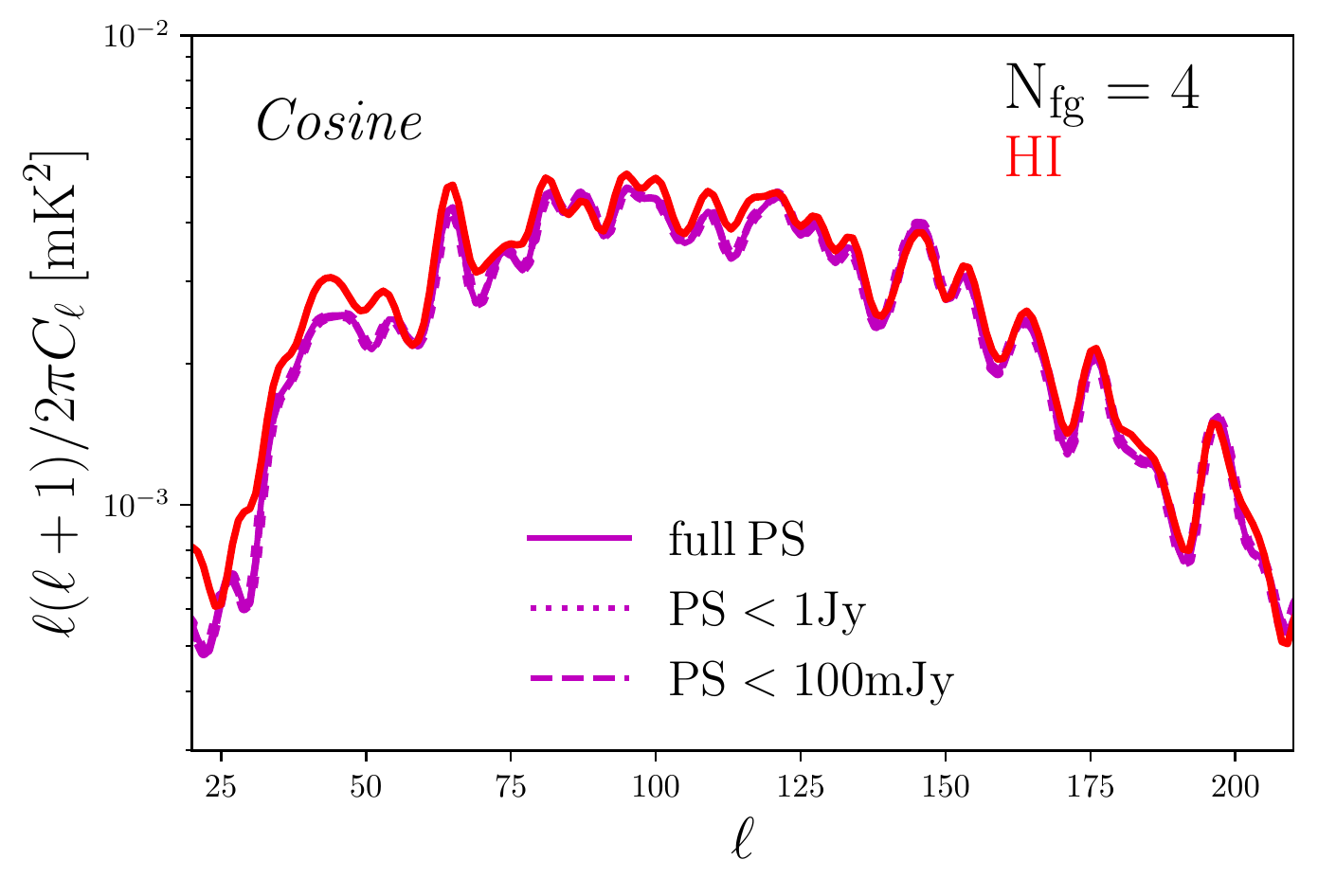}
\includegraphics[width=8cm]{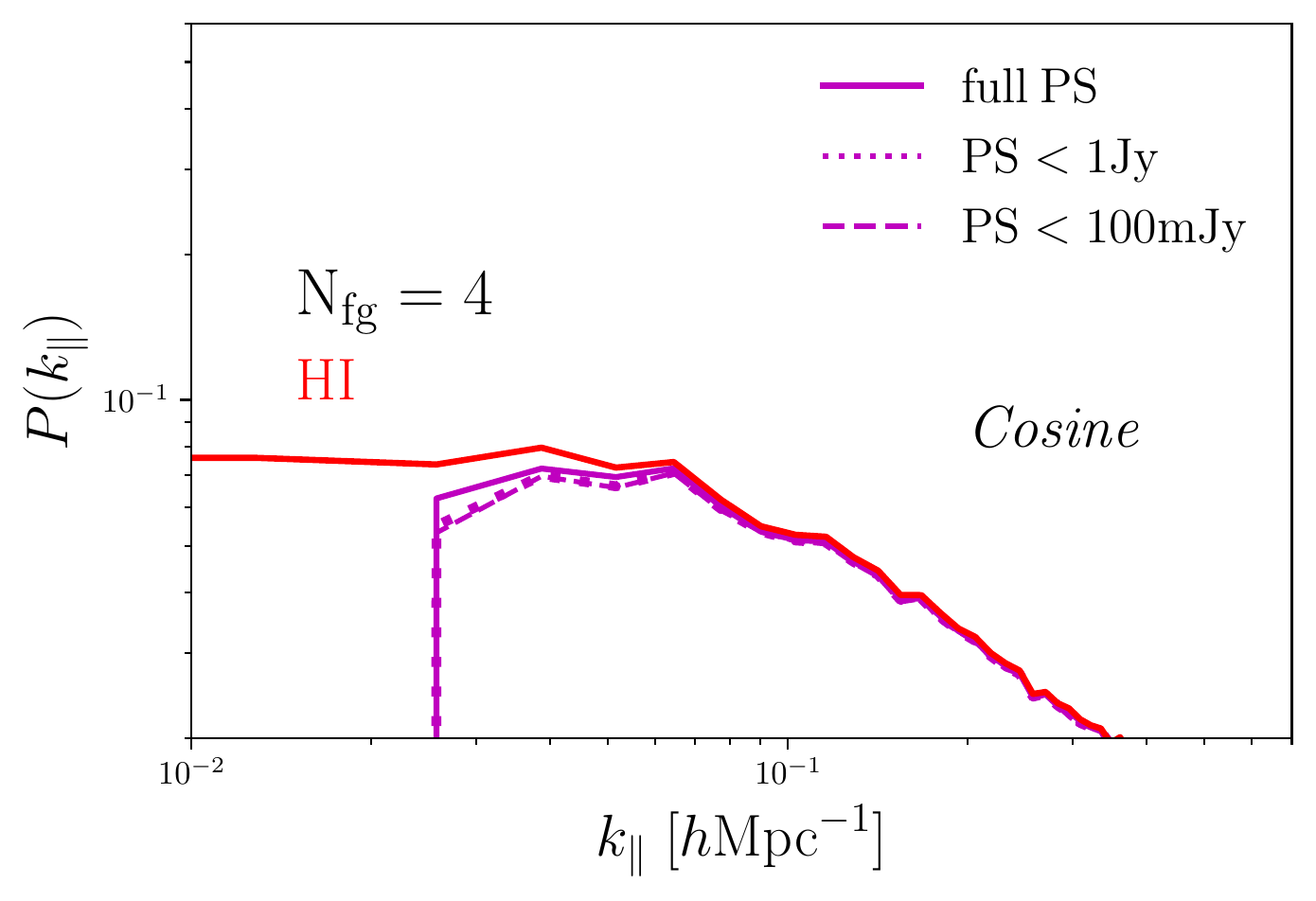}
\caption{The results for the foreground cleaning estimators considered in this work: the angular power spectrum defined in equation~\ref{eq:cl} (top panel) for a given frequency ($950~\mathsf{MHz}$ in this example) and the radial power spectrum of equation~\ref{eq:Pk} (lower panel). We assume a {\it Cosine} beam model for the simulated observations and that the FWHM scales proportionally to $\lambda/D$. Different line-styles correspond to different levels of point source contamination. The solid curves represents the case with no flux cut applied to the catalogue (full PS), PS$ <1 $~Jy is in dotted and PS$ <100$~mJy in dashed. 
We present results using $N_{{\rm fg}}=4$ for the number of removed components. The retrieved signal is compared to the input HI signal plotted in red.}
 \label{fig:clean_cos}
 \end{figure}

\begin{figure}
\includegraphics[width=8cm]{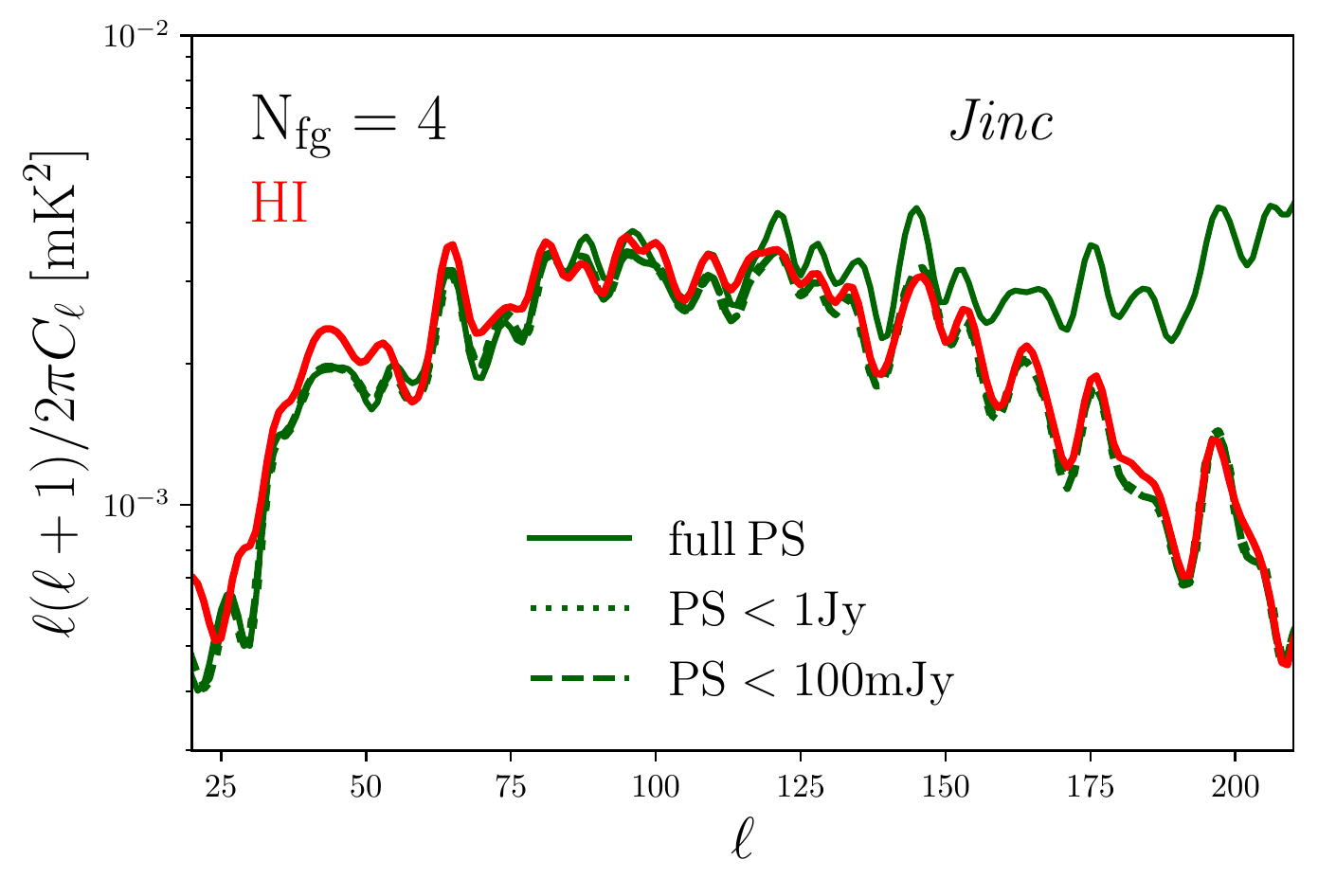}
\includegraphics[width=8cm]{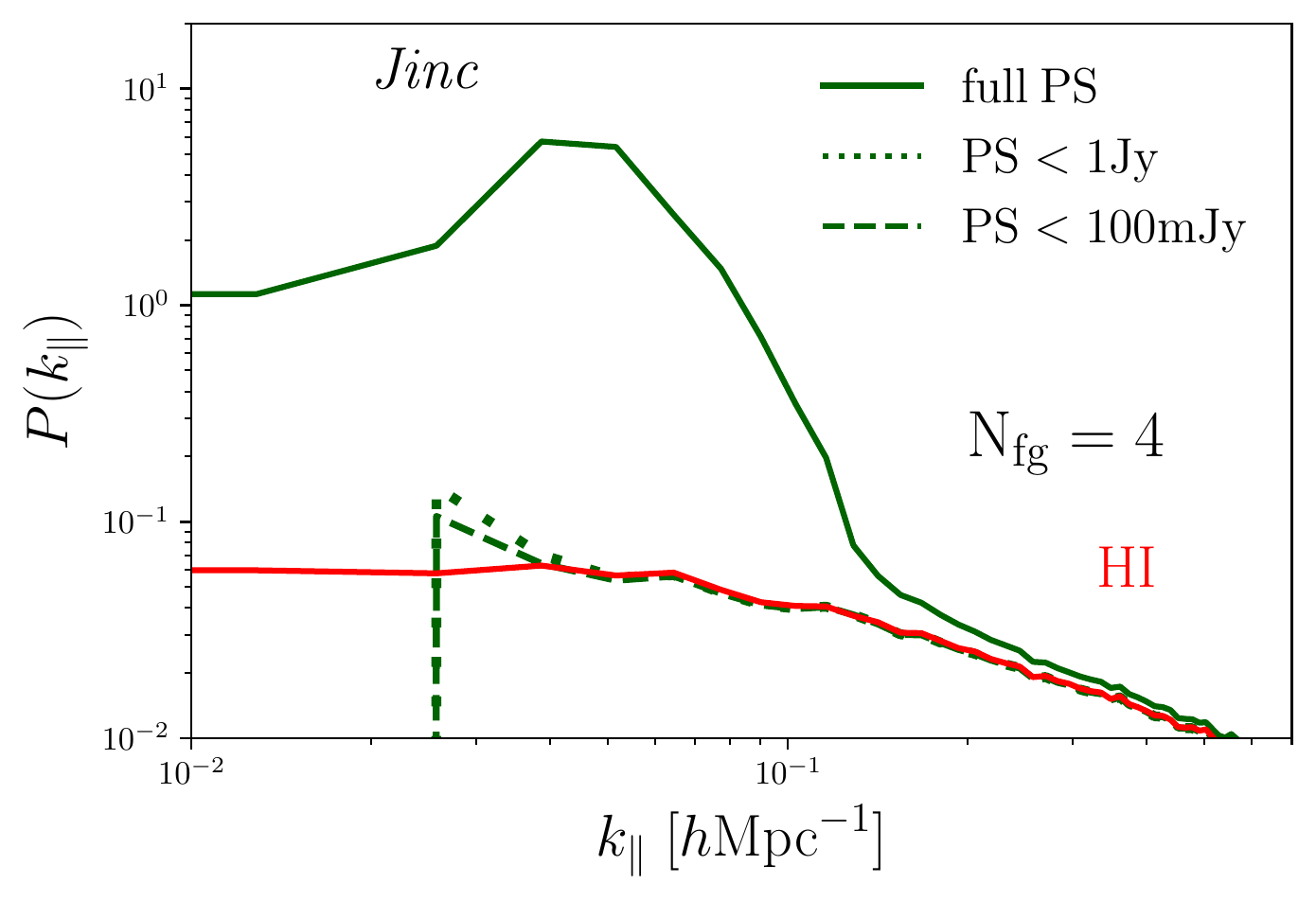}
\caption{Same as figure~\ref{fig:clean_cos}, but for a {\it Jinc} beam model that has stronger sidelobes with respect to the {\it Cosine} beam. The input HI is shown in the red solid curve whilst the foreground cleaned signals for different flux density cuts are shown with different line styles.}
\label{fig:clean_jinc}
 \end{figure}

\begin{figure*}
\includegraphics[width=8cm]{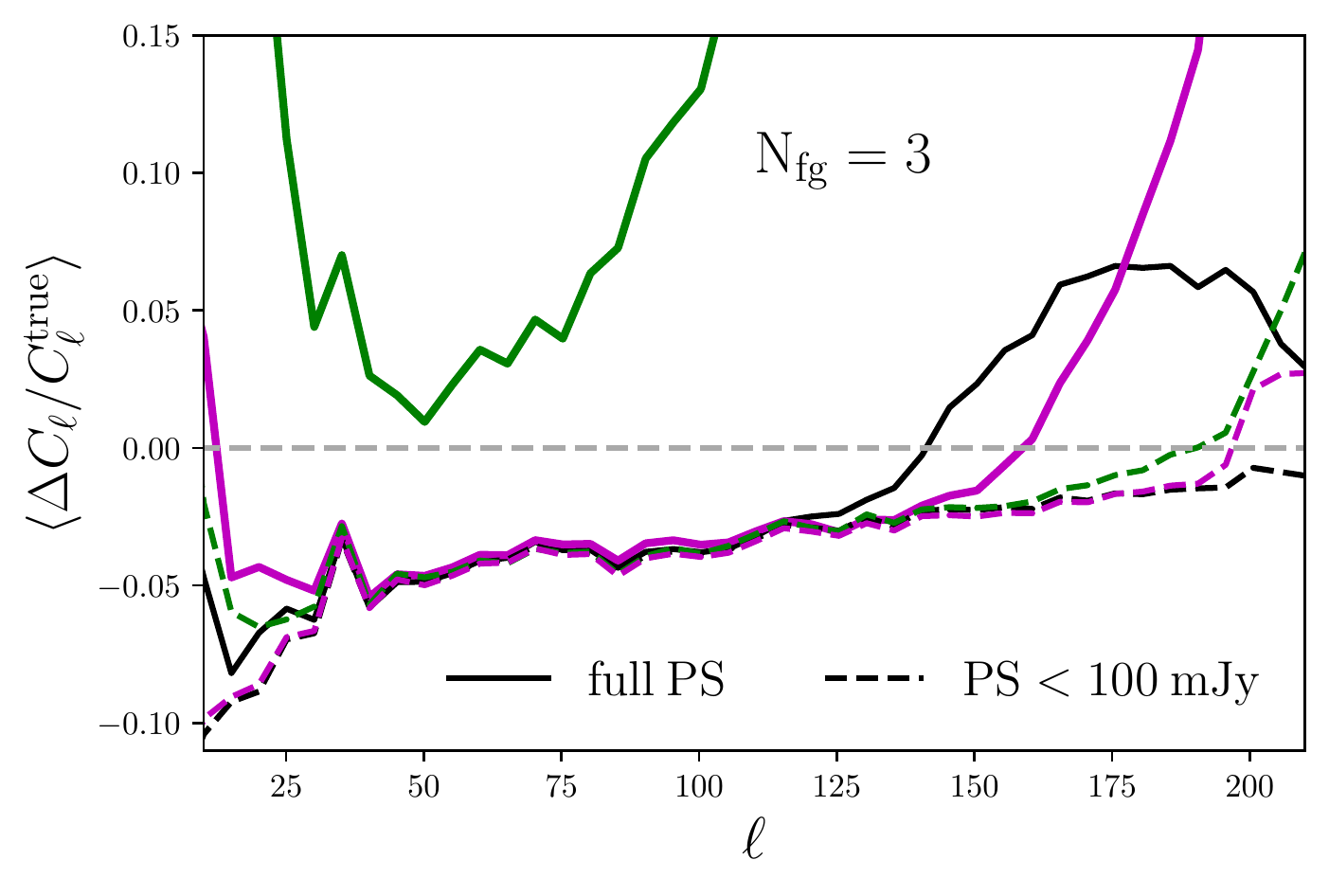}
\includegraphics[width=8cm]{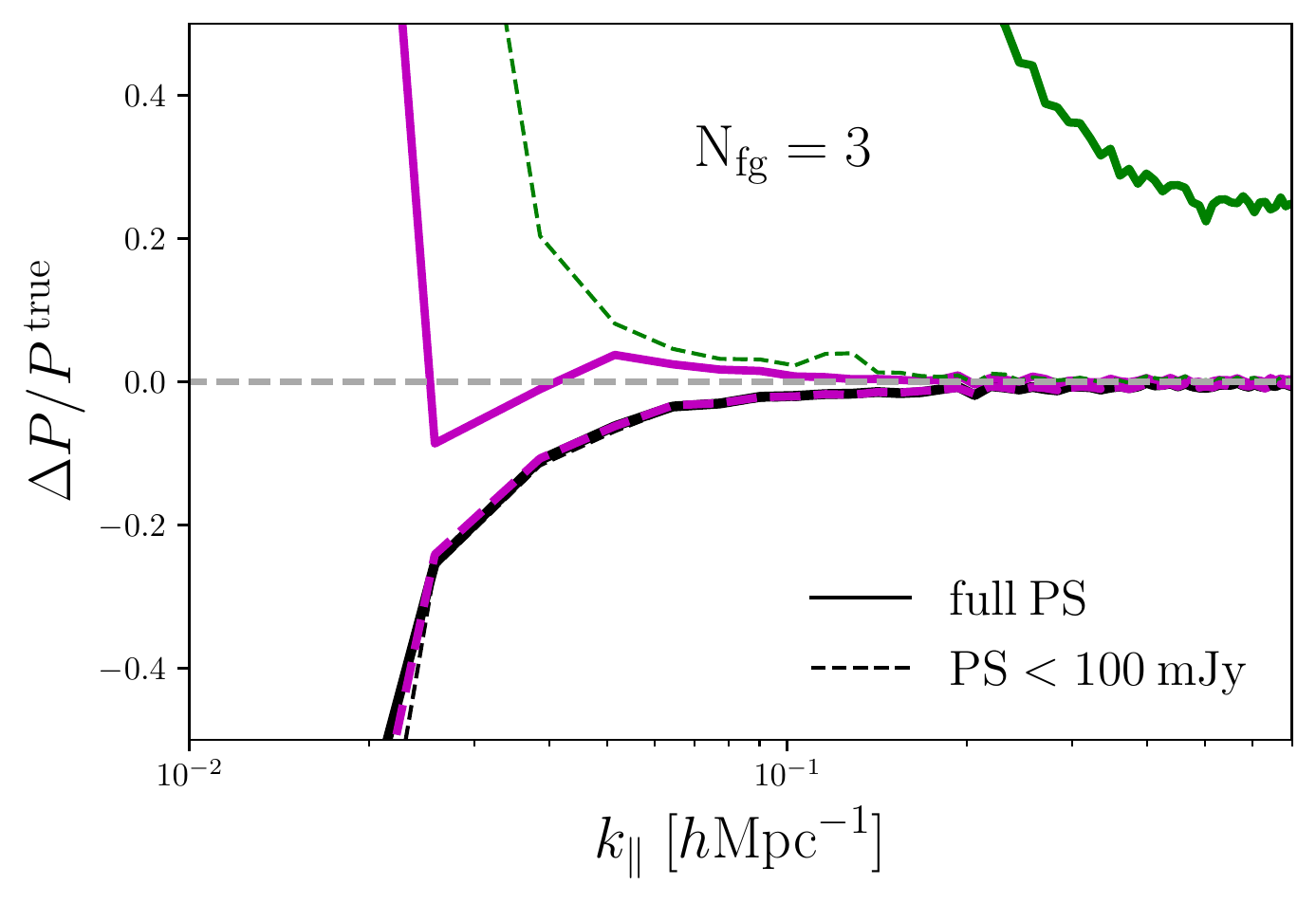}\\
\includegraphics[width=8cm]{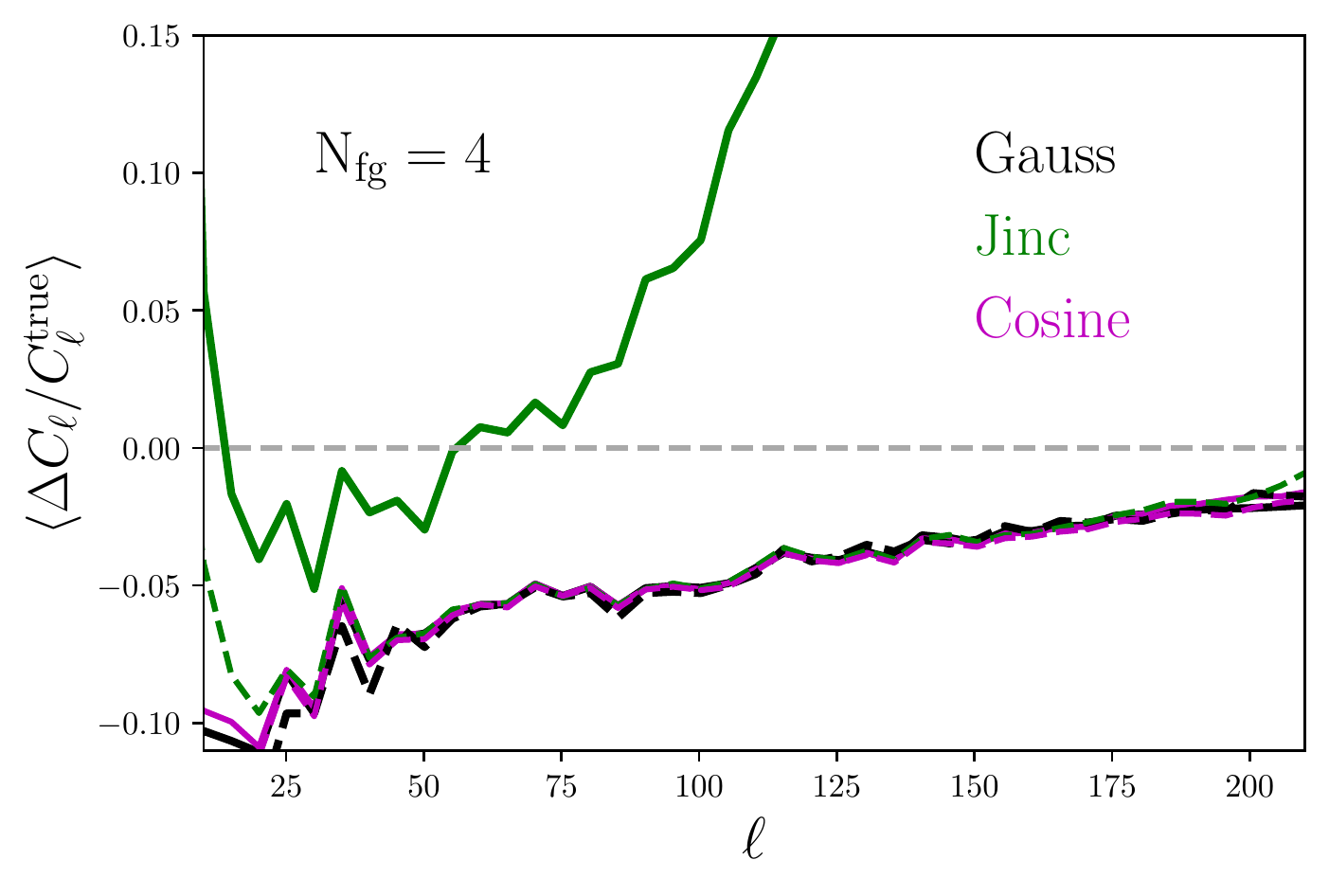}
\includegraphics[width=8cm]{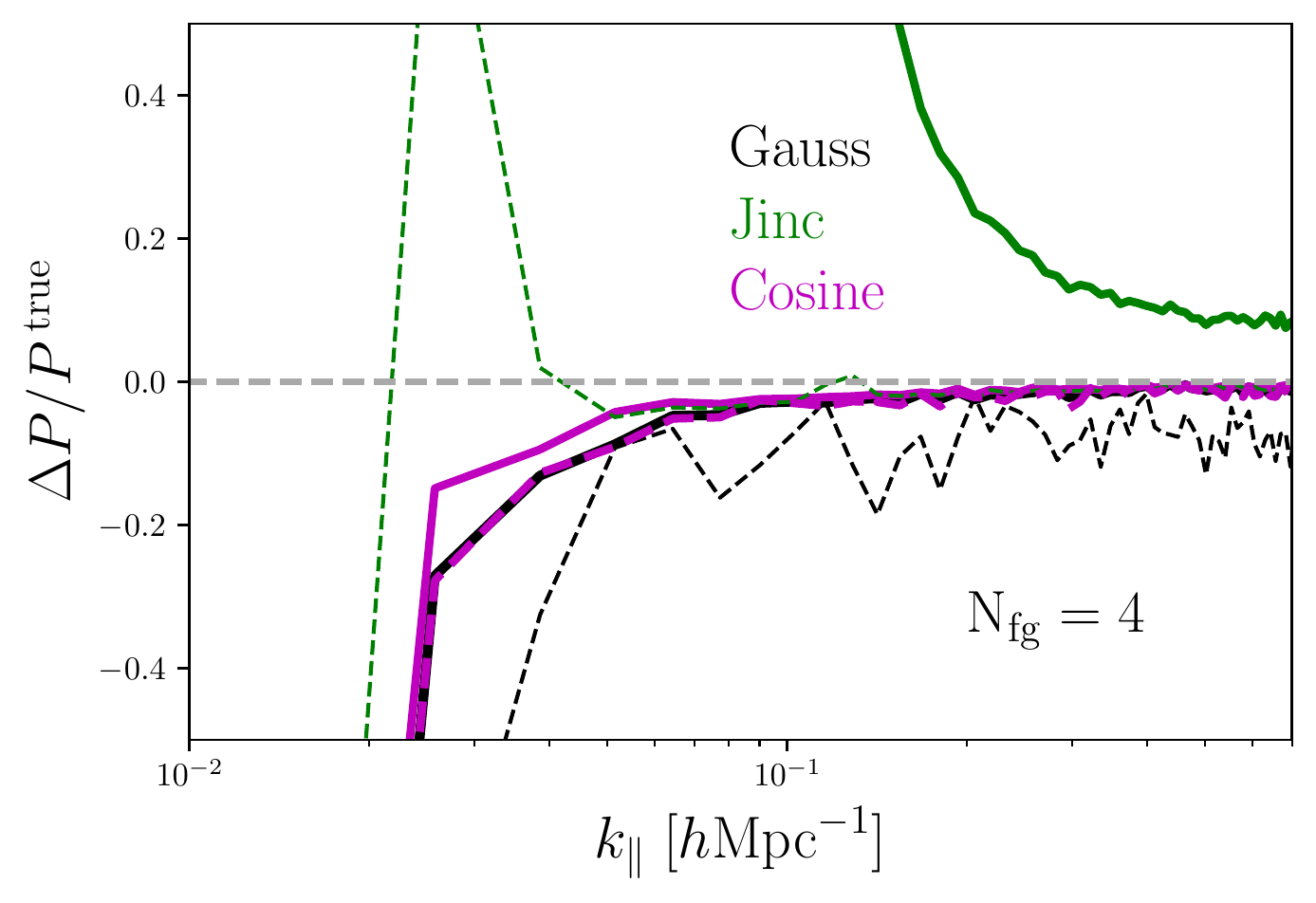}
\caption{A quantitative comparison of the foreground cleaned angular power spectrum (left) and radial power spectrum (right) with the true input signal (see equation~\ref{eq:Dcl} and \ref{eq:Dpk}). Three types of beam models, \textit{Gauss} (black), \textit{Jinc} (green) and \textit{Cosine} (magenta) were used in the beam convolution pipeline. In all cases the FWHM scales simply with $\lambda/D$. The performance of the recovery of the HI signal is evaluated for $N_{\rm fg}=3$ in the top panels and for $N_{\rm fg}=4$ in the bottom panels. The solid curve represents the case in which the full point source catalogue was considered for the simulated maps, whilst the dotted lines are the cases in which a flux cut of PS$<100$~mJy was applied.} 
 \label{fig:fgrm_noripple}
 \end{figure*}

  \begin{figure}
\includegraphics[width=8cm]{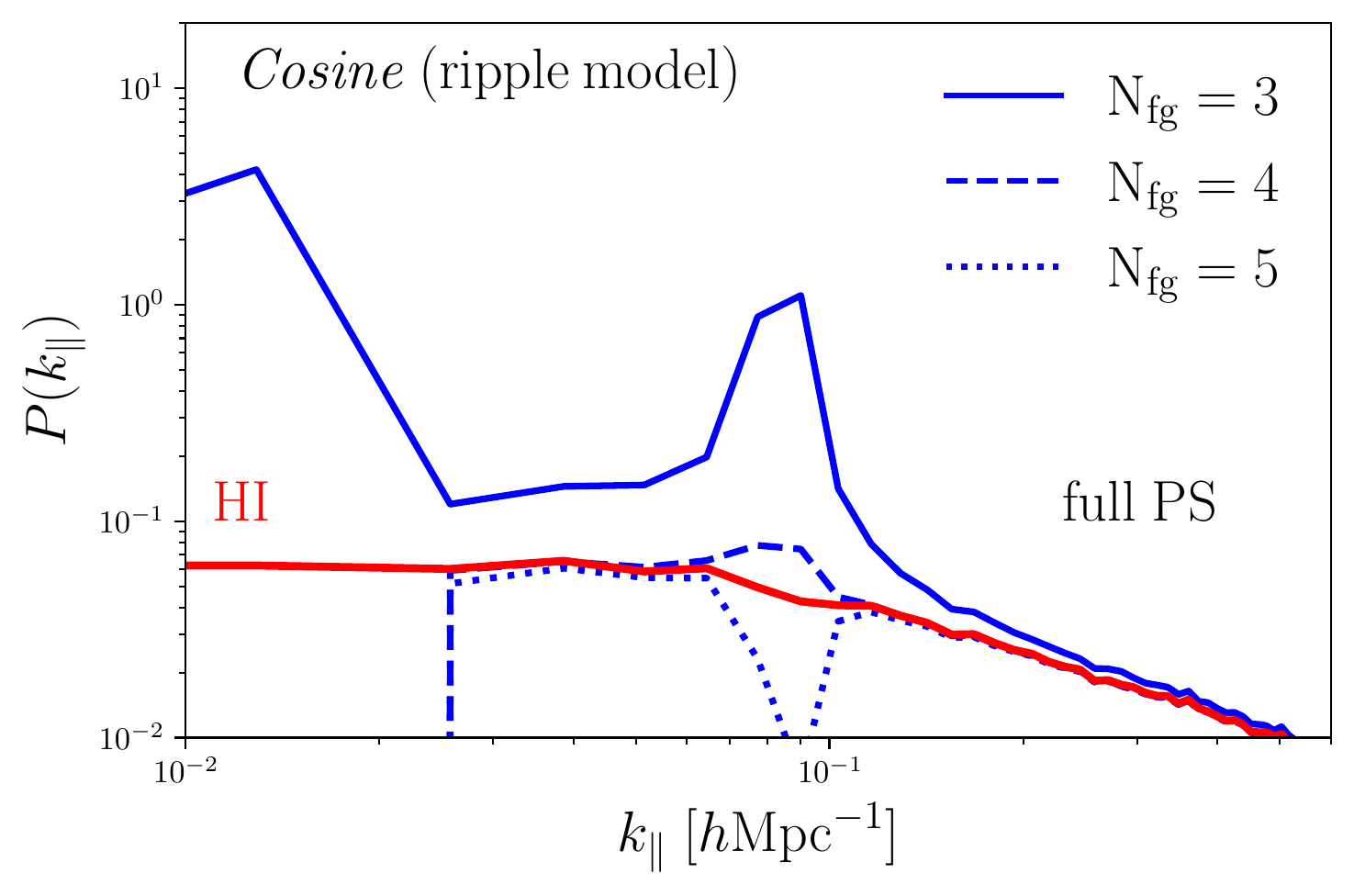}
\includegraphics[width=8cm]{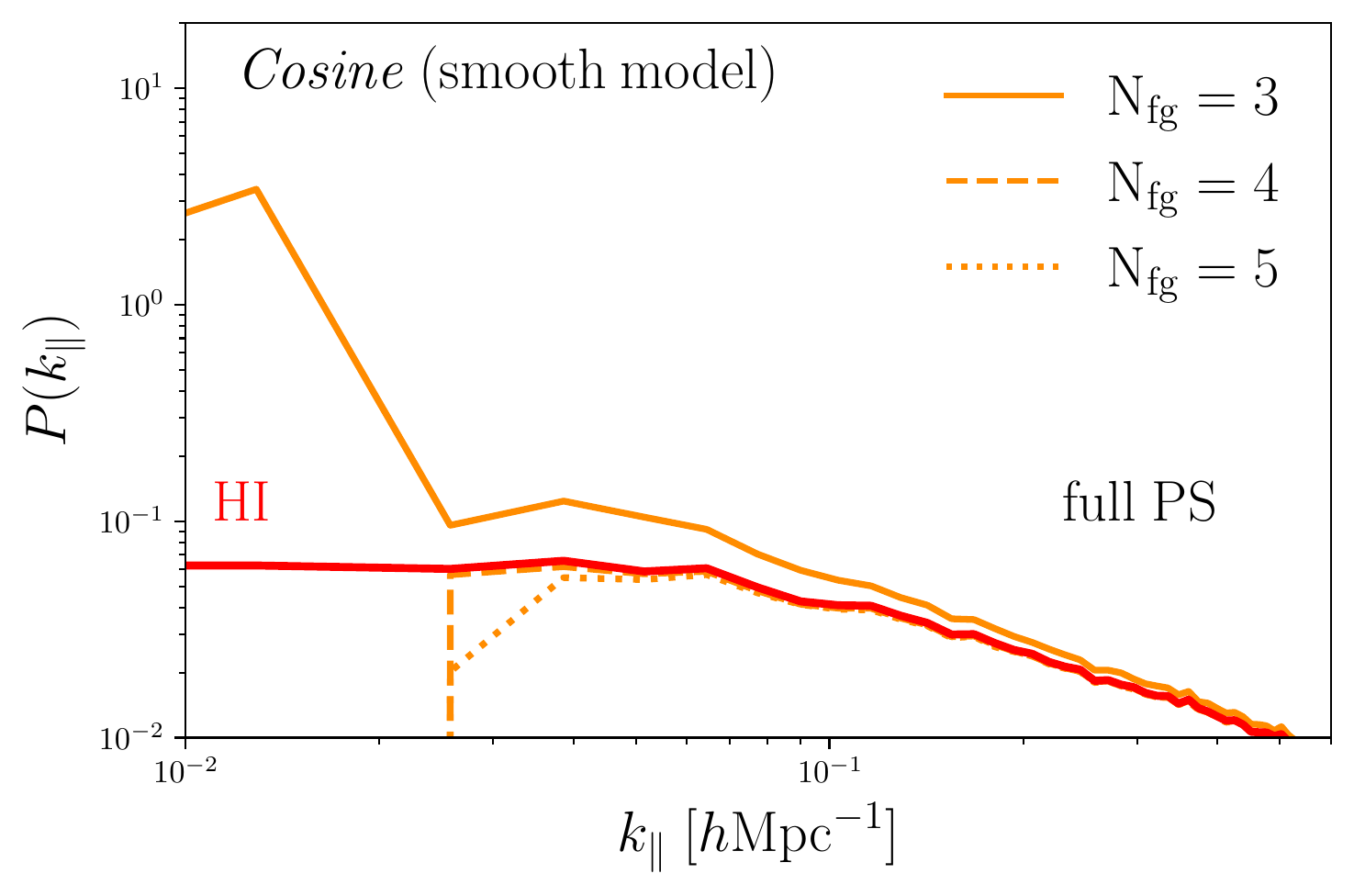}
\caption{Comparison of radial power spectra of the recovered signal for the {\it Cosine} beam case with a FWHM with non-trivial spectral variation: the 
full {\it ripple} model (top panel) or only the polynomial {\it smooth}  model (bottom panel). The input HI is shown in red whilst the reconstructed signals for different number of foreground components removed ($N_{{\rm fg}}$) are shown with different line styles. This results are obtained without removing the strong point sources.}
 \label{fig:ripple_PS}
 \end{figure}

As discussed in section~\ref{sec:beam}, a realistic model for the MeerKAT beam should include sidelobes. 
Since sidelobes should be most problematic in the presence of strong point sources, we show results considering different levels of point source contamination. Our best-case scenario considers that we will be capable of removing point sources with fluxes as low as $100$~mJy at $1.4~\mathsf{GHz}$, but we explore also more pessimistic cases.
Note that, at this stage, we are simply considering that the FWHM of the beam scales proportionally to $\lambda/D$ (see equation~\ref{eq:fwhm}). The more complex models of equation~\ref{eq:fwhm_nu} will be explored later on.

A good level of approximation for the shape of the sidelobes in the case of MeerKAT, is obtained with the cosine-tapered field illumination function, i.e. the {\it Cosine} model. We present in section~\ref{sec:cos_beam} the results obtained using mock sky emission convolved with this type of beam. 
A more pessimistic assumption for the sidelobes is instead presented in section~\ref{sec:jinc_beam}, where we consider the \textit{Jinc} model (see figure \ref{fig:beam_models}). 

\subsubsection{Realistic sidelobes}\label{sec:cos_beam}
We consider here the sky model convolved with a \textit{Cosine} beam and apply the blind cleaning.
An example of the reconstructed angular power spectrum is shown in the upper panel of figure~\ref{fig:clean_cos}. Despite the sidelobes of the \textit{Cosine} beam, the foreground cleaning method has no particular difficulties in reconstructing the HI signal and there is no dependence on the level of point source contamination. The results presented in figure~\ref{fig:clean_cos}  are for $N_{\rm fg}=4$, which we found was the optimal number of components to be subtracted. Rising $N_{\rm fg}$ only worsens the over-cleaning at large scales.

When examining the radial power spectrum (lower panel of figure~\ref{fig:clean_cos}) we find that the overall quality of the cleaning is good for intermediate and small scales. As for the angular power spectrum, we show results for $N_{\rm fg}=4$. Large scales are inevitably over-cleaned since the strong foregrounds are mostly smooth in frequency thus making power at low $k_\parallel$ the first to be subtracted by blind methods. We note that the cleaning, at fixed $N_{\rm fg}$, is more aggressive for lower point source contamination. This is expected since the complexity created by the interaction between sidelobes and strong point sources could counteract signal loss. 
Moreover, we report that, for $N_{\rm fg}=3$, the case with the full point sources still leaves high contamination on large scales. On the other end, as soon as the strongest point sources are not included in the sky model, choosing $N_{{\rm fg}}=5$ leads to severe over-cleaning, subtracting power not only at large scales but at intermediate $k_\parallel$.

\subsubsection{Pessimistic sidelobes}\label{sec:jinc_beam}

We now consider a more pessimistic model for the beam (worse than what is expected for MeerKAT) by convolving the sky with the \textit{Jinc} beam and apply again the blind cleaning. We present again only the results with $N_{\rm fg}=4$.
In the upper panel of figure~\ref{fig:clean_jinc}, we can see that in this case the performance of the cleaning algorithm is poorer: the full point source case has residual contamination already at intermediate scales.
A higher $N_{\rm fg}$ does not ease this contamination and worsens the signal loss at large scales.
We can appreciate even better the role of strong sidelobes and point sources in the cleaning procedure when examining the radial power spectrum (lower panel of figure~\ref{fig:clean_jinc}). We observe a "bump" in the low to mid $k_{\parallel}$ scales for the full point source model, and the HI radial power spectrum  is only recovered at large $k_{\parallel}$. Increasing the number of foreground components $N_{{\rm fg}}$ has a very small effect in removing this feature. Residual foregrounds are gradually removed when point source flux cuts are applied, and the HI signal is recovered at least at intermediate and small scales.

\subsubsection{Quantitative comparison of beam models}\label{sec:recovery}
We compare in figure~\ref{fig:fgrm_noripple} the cleaning results for the {\it Cosine} and {\it Jinc} beam model with the input signal. We add for reference the case without sidelobes, i.e. the {\it Gaussian} model. We show the estimators of equation~\ref{eq:Dcl} and \ref{eq:Dpk}. By construction, $\Delta P/P^{\rm true}(k)$ and $\Delta C_{\ell}/C_{\ell}^{\rm true}$ will be positive if there are still contaminants in the recovered signal, and negative if the cleaning is too aggressive, resulting in signal loss.
We retain the worst and best-case scenario for the point source contamination, considering the full catalogue and point sources with a flux cut of $100$~mJy, respectively.

The complexity of the interaction between point sources and sidelobes appears clearly from the figure: for the {\it Gaussian} case there is little dependence on the level of point source contamination while, for the two models with sidelobes, the cleaning is easier without strong point sources. Note also that, while $N_{{\rm fg}}=3$ is the optimal assumption for the {\it Gaussian} case, the {\it Cosine} and the {\it Jinc} beam requires a higher number of components to be subtracted.
When the \textit{Jinc} beam is convolved with a sky model with no cut on point source flux, the reconstructed signal remains highly contaminated by foregrounds and increasing $N_{{\rm fg}}$ offers little assistance. Nevertheless, when considering faint point sources (PS~$ < 100$~mJy), the angular power spectrum signal is recovered with 10\% precision reaching a few \% at small scales, and it is possible to recover the radial power spectrum, except for low $k_\parallel$.
For the more realistic {\it Cosine} beam, sidelobes seems not to be a major limitation in the recovery of the signal even in the presence of strong point sources, and using the optimal value of $N_{{\rm fg}}=4$, we can recover the HI signal within 10\% precision.

A note of caution is needed: the recovered signal is systematically below the input signal for the optimal value of $N_{{\rm fg}}$, indicating signal loss.
This effect can be, in principle, corrected by constructing a foreground transfer function, generally repeating the cleaning on a set of simulation \citep[e.g.][]{switzer2015,wolz2021}. Although we do not characterize this function in this work, figure~\ref{fig:fgrm_noripple} gives a raw expectation of how the signal loss varies across the number of components to subtract, the level of point source contamination and the primary beam model.

\subsection{Frequency dependent beam}\label{sec:results_freq}

\begin{figure*}
\includegraphics[width=8cm]{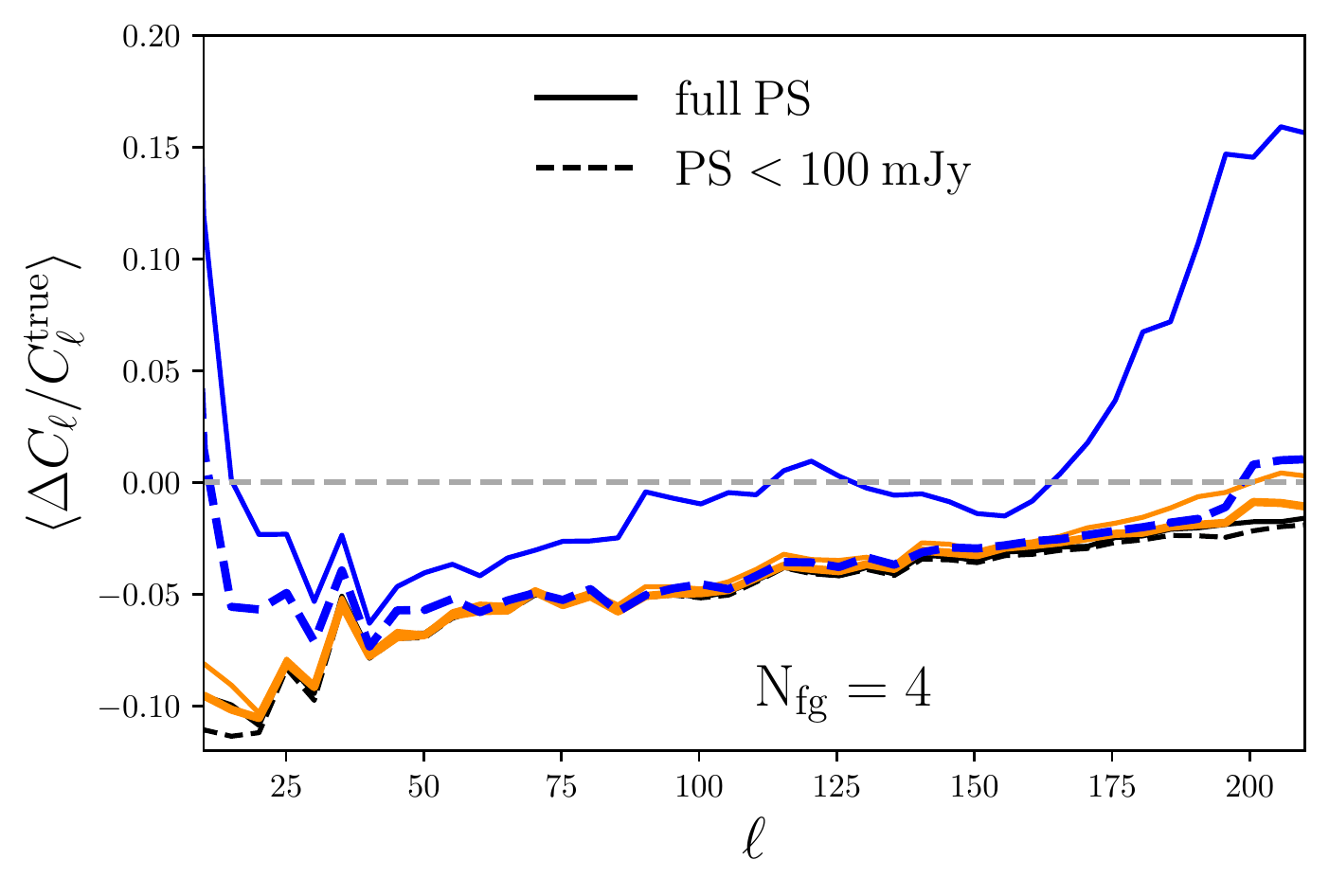}
\includegraphics[width=8cm]{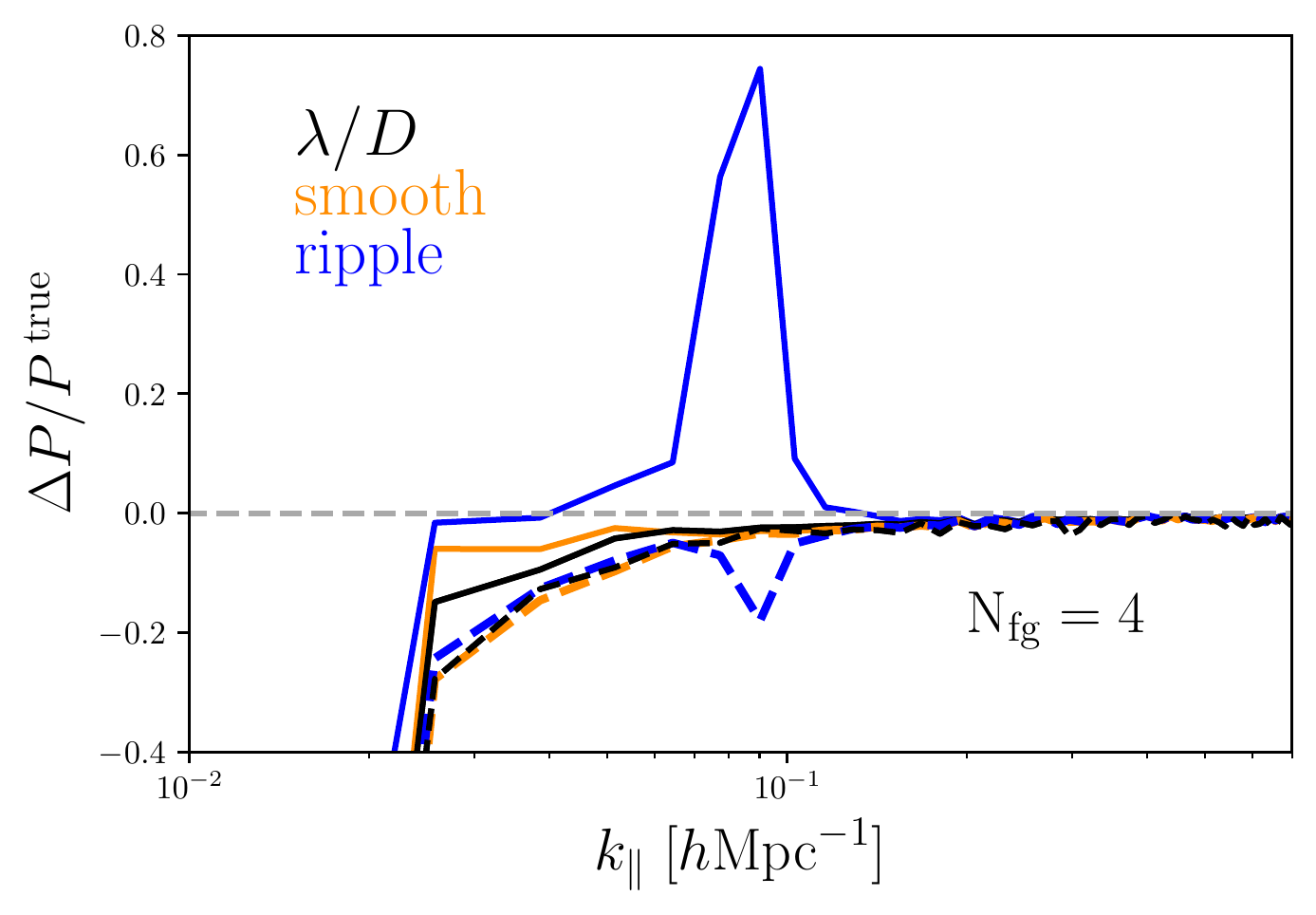}
\caption{A quantitative comparison of the foreground cleaned angular power spectrum (left) and radial power spectrum (right) with the true input signal (see equation~\ref{eq:Dcl} and \ref{eq:Dpk}).
Results for $N_{\rm fg}=4$ are compared for the different FWHM models: $\frac{\lambda}{D}$ in black, the smooth model in orange and the ripple model in blue. Results are shown for the worst and best-case scenario for the point source contamination, considering the full catalogue (solid lines) or only point sources with a flux cut of $100$~mJy (dotted lines), respectively.}

 \label{fig:fgrm_ripple}
 \end{figure*}
 
\begin{figure}
\includegraphics[width=8cm]{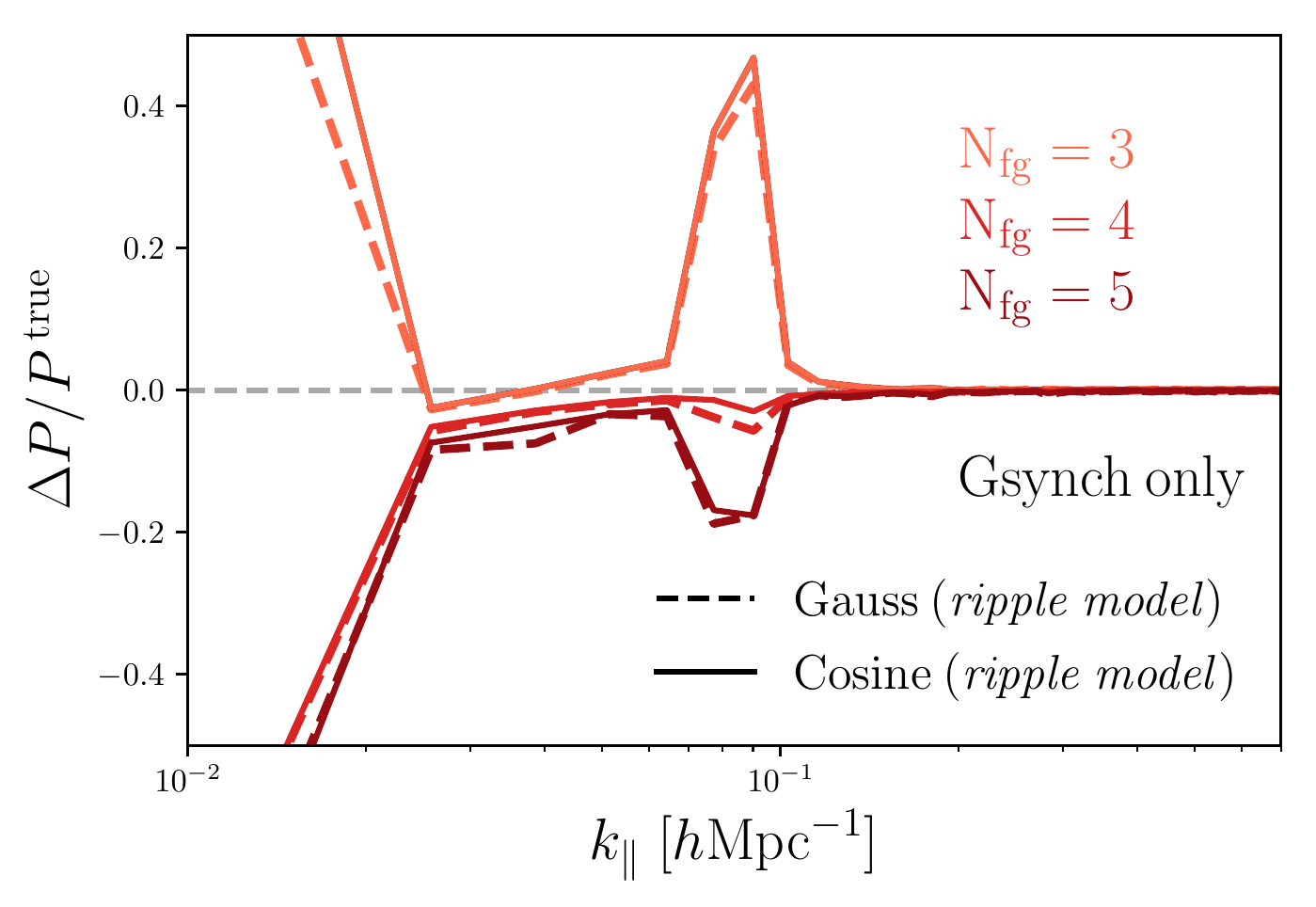}
\caption{The estimator of equation \ref{eq:Dpk} applied after cleaning an input simulation that includes {\it only} synchrotron emission and the signal, convolved with a {\it Cosine} beam  (solid lines) or a {\it Gauss} beam (dashed lines), in both cases assuming the frequency \textit{ripple} model. Different colors show results obtained with different values of $N_{\rm fg}$. Although the effect is similar, note that for $N_{\rm fg}=4$ it is only at few \% level while it reaches 60\% in figure~\ref{fig:fgrm_ripple} in the presence of strong point sources.}

 \label{fig:fgrm_gsynch}
 \end{figure}

Up to now, we have assumed the standard $\lambda/D$ dependence for the FWHM of the beam. What if we relax this hypothesis? We investigate here how our conclusions change in the presence of either the {\it smooth} or the {\it ripple} models presented in section~\ref{sec:beam}. We note again that this frequency dependence cannot be simply absorbed into the bandpass calibration. Throughout this section, we use our most realistic model for the MeerKAT beam: the \textit{Cosine} beam model.

Being a function of frequency, the effect of a non trivial FWHM is mostly visible in the reconstructed radial power spectrum.  We present in the upper panel of figure~\ref{fig:ripple_PS} results without subtracting any point sources from the foreground models. We note here, and elaborate on later, that the trends are similar for the cases with lower contamination, although the magnitude of the effect is smaller.

The radial power spectrum for the {\it ripple} model shows a feature around $\rm{k}_{\parallel}$ = 0.1 $\rm{h Mpc}^{-1}$  whose position depends on the period of the oscillation of the FWHM, in this case $\rm{T=} 20~\mathsf{MHz}$. We find that the amplitude of the feature can be reduced by making the cleaning more aggressive, although we quickly start to see a strong depletion of the signal at the same $\rm{k}_{\parallel}$. The HI signal is best recovered with $N_{{\rm fg}}=4$. 
In the presence of the {\it smooth} model (lower panel of figure~\ref{fig:ripple_PS}) the feature caused by the ripple disappears and the performances of the cleaning are similar to what is seen for the {\it Cosine} beam with the standard FWHM in figure~\ref{fig:clean_cos}.

\subsubsection{Quantitative comparison of spectral models}\label{sec:DeltaFWHM}
In figure~\ref{fig:fgrm_ripple}, we evaluate quantitatively the cleaning at $N_{{\rm fg}}=4$ and present the results not only for the full point source catalogue, but also for the flux cut at $100$~mJy.
The HI angular power spectrum is recovered within 10\% precision for the {\it smooth} model, reaching the same performances of the standard $\lambda/D$ case, with both strong or low point source emission. In the {\it ripple} case instead, if no point source cut is applied, the cleaning methods struggle in recovering the small scales.
Moreover, while the radial power spectrum can be successfully recovered for the standard $\lambda/D$ model and the {\it smooth} model, the cleaning method is 60\% off at $\rm{k}_{\parallel}\sim 0.1$ h$\rm{Mpc}^{-1}$ for the {\it ripple} case.
For this latter case, the removal of the bright point sources transforms the feature of residual contamination in a signal loss at the same scales reaching up to 20\%.

\subsubsection{Impact of diffuse emission}\label{sec:diffuse}
Up to now we have focused our discussion on the interaction between the structure of the beam and the level of point source contamination. We remark that our foreground model includes not only point sources but also free-free and synchrotron emission. A legitimate question is then how much the cleaning performance is impacted by the {\it ripple} model in the presence of a diffuse foreground component. We consider a simplified foreground emission comprising only Galactic synchrotron and report in figure~\ref{fig:fgrm_gsynch} the results for the HI radial power spectrum. 
Similarly to figure \ref{fig:ripple_PS} 
and \ref{fig:fgrm_ripple}, the cleaning struggles around the scale of the FWHM frequency oscillation, although its impact is smaller. Interestingly, the same exercise performed with the {\it Gaussian} beam convolution yields similar results,  underlying that the effect of the persistent {\it ripple} is important also when no sidelobes are present. Note that if the foregrounds were completely smooth across the sky, no effect would be observed.

\subsection{Applying re-smoothing}\label{sec:resmooth}

\begin{figure*}
 \includegraphics[width=18cm]{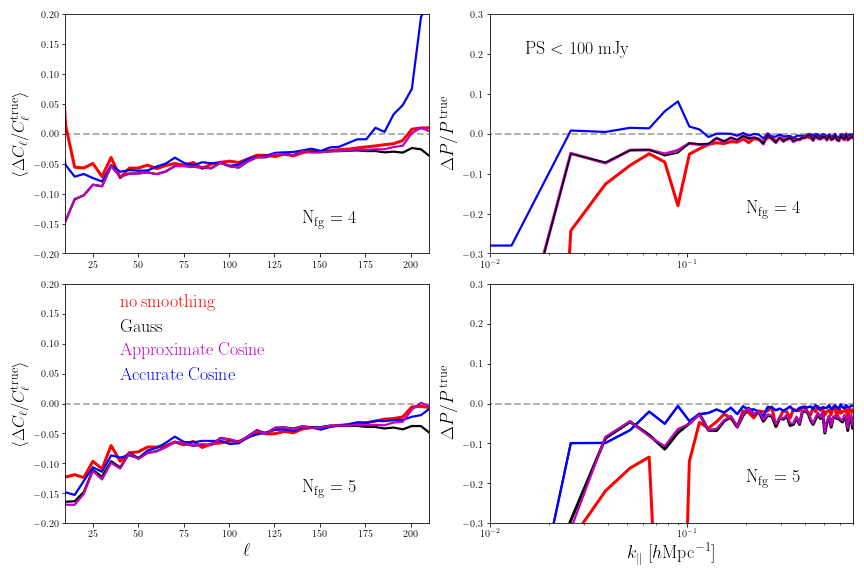}
\caption{We compare the signal reconstruction performances for the {\it Cosine} beam case with a FWHM evolving with the {\it ripple model}, when our standard cleaning procedure is applied and maps are not smoothed (red), with the one obtained after performing a re-smoothing with a Gaussian (black), an approximated cosine (magenta) or a accurate cosine (blue) beam (see text for details), at fixed $N_{\rm fg}$. We show results for the case with only point sources fainter than $100$ mJy, for both the angular power spectrum (left) and the radial power spectrum (right).}
 \label{fig:fgrm_smoothing}
 \end{figure*}

To alleviate the effect of the frequency {\it ripple} in the cleaning process, that, as seen in section~\ref{sec:results_freq}, compromises the recovery of the radial power spectrum, one could attempt a re-smoothing of the maps to a common resolution. Re-smoothing is a standard technique adopted in Intensity Mapping to prevent systematics and artifacts in the data: all maps are de-convolved and re-smoothed to a common angular resolution, slightly lower than the worst one in the data (see for example \citet{anderson2018}).
In other words, a map measured at frequency $\nu$ and thus convolved with a telescope beam with a certain FWHM can be adjusted to a (lower) frequency resolution given by the FWHM of the beam at a lower frequency $\nu_0$. This is achieved in spherical harmonics space via a simple multiplication  of the $a^{\rm mes}_{\ell m}(\nu)$ of the measured map with the ratio of the spherical harmonic representations $b_\ell(\nu_0)$ and $b_\ell(\nu)$ of the two beams 

\begin{equation}\label{eq:alm_res}
    a^{\rm rs}_{\ell m}(\nu)=a^{\rm mes}_{\ell m}(\nu) \frac{b_\ell (\nu_0)}{b_\ell(\nu)}.
\end{equation}
Although this procedure is exact for a full sky map with known beam and no noise, its application to smaller sky patches requires some precautions due to numerical instabilities, which we briefly introduce here and discuss in more detail in appendix~\ref{app:resmooth}. 
\begin{itemize}
    \item[i)] We start by {\it apodizing} our mask (e.g. smoothing the edges) to have less "ringing" entering the computation of the $a^{\rm mes}_{\ell m}(\nu)$;
    \item[ii)] assuming a certain beam model, we compute the $b_\ell(\nu)$ of the frequency of interest and the $b_\ell(\nu_0)$ of the common frequency we want to resmooth to;
    \item[iii)] we regularize the ratio $b_\ell (\nu_0)/b_\ell(\nu)$ imposing a cut-off $\ell_{\rm cut}$ at small scales;
    \item[iv)] we apply equation~\ref{eq:alm_res} and obtain a map from $a^{\rm rs}_{\ell m}$;
    \item[v)] we enforce a conservative mask on the re-smoothed map in order to avoid the inclusion of pixels at the edges of the old mask.
\end{itemize}
We apply this procedure to both the pessimistic case, where no point source flux cut has been applied (full PS), and the best-case scenario where their emission has been kept lower than $100$mJy.  
All our maps have been initially convolved with the {\it Cosine} beam with a peculiar frequency dependence (the {\it ripple} of equation~\ref{eq:fwhm_nu}). 
The deconvolution (dividing by $b_\ell(\nu)$) should be done with a beam that we believe closest to the real one.
We then consider three distinct cases for the deconvolution step, assuming increasing knowledge on the beam.
\begin{itemize}
    \item[1.] A {\it Gaussian} beam with a FWHM scaling proportionally to $\lambda/D$ as in equation~\ref{eq:fwhm}. This is the simplest assumption;
    \item[2.] A {\it Cosine} beam. Since the presence of sidelobes is a known feature of the MeerKAT beam, we can suppose that we can describe them with this model. We instead assume no knowledge of the frequency behavior of the FWHM and rely on the $\lambda/D$ approximation. We call this case {\it approximate Cosine};
     \item[3.] A {\it Cosine} beam with ripple model, assuming exact knowledge of the beam sidelobes and the frequency dependence. We call this case {\it accurate Cosine}. 
     Note that, even in this last case where we deconvolve using the same $b_\ell$ used for the original convolution, the cancellation will not be perfect due to the mask and the noise (although the noise effect should be negligible on the foreground map).
\end{itemize}
The subsequent smoothing (convolution) to a common resolution (using $b_\ell(\nu_0)$), could be done with different beam shapes. A common solution is to use a Gaussian. In this work, we use instead the same beam that was used in the deconvolution step above.

In figure~\ref{fig:fgrm_smoothing}, we compare the performance of the cleaning algorithm using maps de-convolved/re-smoothed with the three different procedures. We present both the angular power spectrum and the radial power spectrum for $N_{\rm fg}=4$ and 5. The reference, for comparison, is the case discussed in section~\ref{sec:results_freq}.
We present, for simplicity, only the case with low point source contamination. This is a conservative choice since we see even more improvement due to the re-smoothing procedure in the case of strong point sources.
Note that we apply exactly the same procedure to the HI only signal so that we can make a fair comparison that should be mostly dependent on the cleaning and not on extra effects from the power spectrum calculation itself (such as the window function).

When we subtract $N_{\rm fg}=4$ components, the $C_\ell$ are similarly reconstructed for all cases within a $\sim 5\%$ precision for most scales, although some residual small-scale contamination is found for the {\it accurate  Cosine} re-smoothing. This residual contamination disappears for $N_{\rm fg}=5$, and we find agreement between the different procedures. We note again the angular power spectrum diagnostic is always negative, uncovering a systematic signal loss.

More interesting for our purposes, is the effect on the  radial power spectrum. The {\it accurate  Cosine} re-smoothing 
deals quite efficiently with the effect of the ripple, reaching a precision always better than 10\% for a large range of $k_\parallel$, for both $N_{\rm fg}=4$ and 5.
The re-smoothing procedure with the {\it Gaussian} or {\it approximate Cosine} beam erases the presence of the ripple and reconstructs the $P(k_\parallel)$ much better than reference case.

\medskip 
In summary, figure~\ref{fig:fgrm_smoothing} suggests that, in the presence of non trivial beam frequency dependent effects, an accurate knowledge of the beam would allow a re-smoothing procedure that should improve the quality of the foreground cleaning in the radial direction and alleviate signal loss. A less accurate re-smoothing would also be enough to retrieve the signal with good accuracy. 
Probably the best approach will be to include the beam in the mapmaking process. Something that will be computationally heavy and we would like to explore in future work. Still, we will never know the beam exactly and a smoothing kernel will always be useful. The choice of smoothing kernel should be validated against simulations in order to quantify any effect on the signal power spectrum.

\section{Conclusions}\label{sec:conclusions}
The SKA precursor MeerKAT telescope in South Africa is a precious test ground for HI Intensity Mapping techniques and has the potential to provide competitive constraints on cosmological observables using the single dish mode \citep{santos2017}.
Component separation techniques such as PCA and FastICA are a vital step to extract the HI signal from the bright foreground emission. These techniques, although widely and successfully tested in literature, have been applied mostly under the assumption of a Gaussian primary beam.
Focusing on MeerKAT characteristics, in this work we presented a detailed analysis of the impact of a more realistic beam on the recovery of the signal.
We explored both the presence of sidelobes and of frequency dependent effects on the FWHM of the beam.
Our sky model includes a new realistic point source full-sky catalogue constructed from the combination of NVSS and SUMSS data and the S$^3$ simulations, which allows to test the effect of strong point sources on the signal.  

We simulated MeerKLASS-like instrumental noise levels and selected a sky patch in the southern hemisphere covering $\sim 9\%$ of the sky, avoiding strong contamination from the Galactic plane.

We analyzed first the effect of sidelobes in presence of strong point sources. Cleaning is severely impacted  by the large sidelobes of the {\it Jinc} beam model and residual foreground contamination improved with the subtraction of a larger number of components (up to $N_{\rm fg}=5$) only if the strongest sources are removed. Fortunately, the {\it Cosine} model, which should be a more realistic description for the MeerKAT beam, shows a good performance if 4 modes are subtracted. Note that $N_{\rm fg}=3$ is enough instead for the {\it Gaussian} beam case. 
We recovered the HI signal within 10\% on large angular scales, while reaching few \% accuracy at small scales. We find, however, that the foreground subtraction always tends to over-clean the signal. 
The HI radial power spectrum reconstruction is good at intermediate and small radial scales but fails at low $k_\parallel$ as expected, due to the removal of the smooth foreground contamination. 
This effect at large scales is known, for example, to bias primordial non-Gaussianity studies \citep[e.g.][]{cunnington2020a}. We found that the performances of the
cleaning at low $k_\parallel$ depend both on the level of point source contamination and on the amplitude of the beam sidelobes.

We then explored the effect of a realistic frequency dependent beam on the cleaning efficiency. In the literature, the beam FWHM is generally assumed to follow a simple scaling ($\lambda/D$). Following the fits of \citet{asad2019}, we constructed two models for the FWHM of increasing complexity using the realistic {\it Cosine} beam (see figure~\ref{fig:fwhmVsfreq}): a polynomial in frequency (i.e the {\it smooth} model) and a frequency oscillation on top of this latter (i.e the {\it ripple} model). 
We found that the {\it smooth} model does not impact the performance of the cleaning procedure.
On the contrary, the coupling of the foreground, and in particular of the strong point sources, with the {\it ripple} feature, affected the cleaning more strongly. The effect is more visible for the radial power spectrum, where the $k_\parallel$ scales corresponding to the frequency of the ripple are biased. It is interesting to note that this effect is also present when only the Galactic synchrotron is included, showing that the contamination will still be present at some level even after aggressive point source removal. Moreover, this contamination also comes from the frequency fluctuations in the main lobe of the beam and not only the sidelobes.
This type of oscillatory behavior of the FWHM of the beam, if not recognized and treated, could contaminate the cosmological analysis, such as the reconstruction of the imprint of the Baryon Acoustic Oscillations on the 21~cm power spectrum. 
Nevertheless, we found that the effect of a beam {\it ripple} can be greatly reduced if the maps are re-smoothed to a common resolution. Overall, the outcome of considering such a realistic beam is positive: sidelobes should not be an issue even when considering strong point sources and the frequency effects seem well localized in $k$ space and can be improved through deconvolution. This is good news for large cosmology surveys with MeerKAT and the SKA. Further studies with more sophisticated foreground cleaning methods might improve this outcome.

\section*{acknowledgements}
The authors would like to thank Khan Asad for his valuable help with the MeerKAT beam modelling and the anonymous referee for the help in improving the clarity of this paper. 
SDM thanks Yichao-Li and Jingying Wang for the useful discussions. SDM would also like to thank Eric Switzer for his valuable comments which guided us in improving the paper.
MS would like to thank Davide Poletti, Isabella Carucci, Giulio Fabbian and Laura Wolz for
their constructive suggestions during the development of this research work.
SDM and MGS acknowledge support from the South African Square Kilometre Array Project 
and National Research Foundation (Grant No. 84156).
We thank the Centre for High Performance Computing (CHPC) for availing their computing resources. SDM acknowledge support of the South African Department of Higher Education and Training (DHET) through the new Generation of Academics Progamme (nGAP).
MS acknowledge funding from the INAF PRIN-SKA 2017 project 1.05.01.88.04 (FORECaST), from the Italian Ministry of Foreign Affairs and
International Cooperation (MAECI Grant Number ZA18GR02) and the South African
Department of Science and Technology's National Research Foundation (DST-NRF
Grant Number 113121) as part of the ISARP RADIOSKY2020 Joint Research Scheme.

\section*{Data availability}
No new data were generated or analyzed in support of this research.




\bibliographystyle{mnras}
\bibliography{biblio} 


\appendix
\section{Point source catalogue extension to polarization}\label{app:ps_cat_P}
It is well know that, due to instrumental imperfections, the polarized sky can leak into the intensity mapping signal. Polarization leakeage can be particularly challenging for foreground cleaning algorithm since its spectral behaviour is not smooth due to Faraday rotation \citep[e.g.][]{spinelli2018}. At radio frequencies, the two main polarized foregrounds are the diffuse Galactic synchrotron and the polarized signal from point sources. Simulations of the former and its impact on intensity mapping has been investigated in \citet{alonso2014} (see also in \citealt{shaw2015} and more recent \citealt{carucci2020,cunnington2020b}), while the latter contaminant remains quite unexplored. 

In this appendix we extend the intensity catalogue presented in section~\ref{sec:ps} to include also polarization. An accurate discussion of the impact of our polarized simulation for IM is ongoing and will be presented in a future work. Note that other simulated polarized source catalogues are also available in the literature \citep[e.g][]{bonaldi2018}.
Our catalogue is the result of the combination of two surveys (NVSS and SUMMS) and the S$^3$ simulations. Neither SUMMS or S$^3$ contain information on source polarization so we use the NVSS polarized flux density $|P|$ data and statistically extrapolate them to our simulated sources. We first
bin the NVSS flux densities into five bins, to have a reasonable amount signal to noise ratio.
We then obtain the polarization fraction for every NVSS source in a bin using $\Pi=|P|/S$ and we compute the mean polarization fraction $\bar{\Pi}_{i}$ and its standard deviation $ \sigma_{\rm {\Pi}_{i}})$ for the $i$-th bin. 
We then divide all the sources in our catalogue in the same five bins and assign to each of them a polarization fraction drawn from a Gaussian distribution $\mathcal{N}(\bar{\Pi}_{i}, \sigma_{\rm {\Pi}_{i}})$.
Every source in the catalogue now has an assigned polarized flux value $|P|=\Pi S$.

\begin{figure}
    \centering
    \includegraphics[width=8cm]{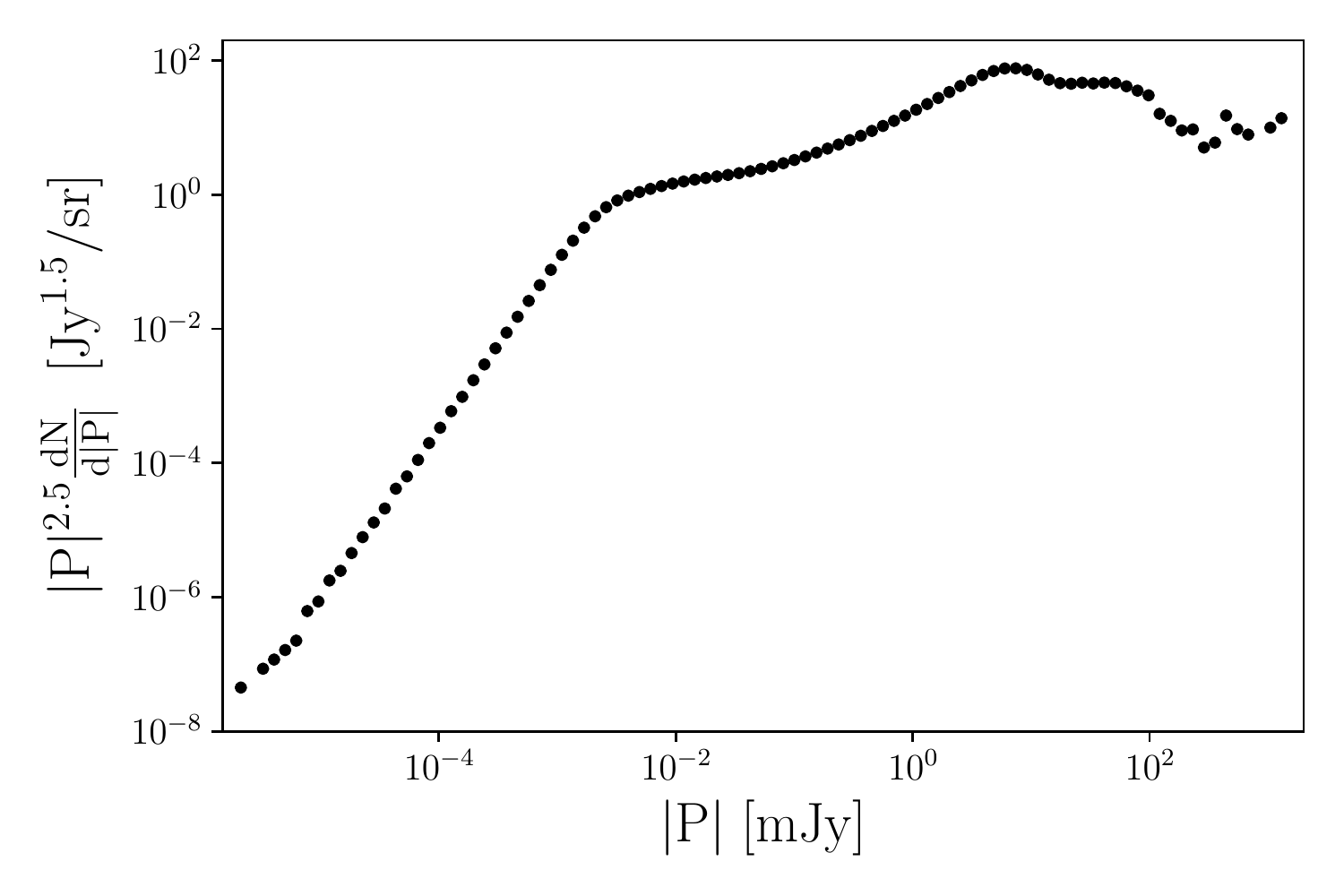}
    \caption{The normalized intrinsic polarized source counts of our catalogue at 1.4GHz. 
    Polarization fractions are computed from NVSS data and extrapolated to all sources.}
    \label{fig:pol_sc}
\end{figure}

To construct Stokes $Q$ and $U$ maps from the catalogue we are still lacking an important information:
the parameters $Q$ and $U$ are related to $|P|$ via
\begin{equation}\label{eq:QU}
\centering
 Q + iU = |P| \ e^{2i\phi},
\end{equation}
where $\phi$ is the polarisation angle given by $\phi = \phi_{0} + \psi \lambda^{2}$. $\phi_{0}$ is the intrinsic polarisation angle that is rotated by the Faraday depth $\psi$ (also called rotation measure), proportional to $\lambda^2$. 
For our catalogue, we assume that the intrinsic polarisation angle can be drawn from a Uniform distribution $\phi_{0} \sim \mathcal{U}(0^{\circ}, 360^{\circ})$. For the Faraday depth $\psi$ instead, we follow \citet{nunhokee2017}, that, motivated by the finding of \citet{taylor2009}, assume $\psi$ values as Gaussian distributed around a mean value of $5.6 {\rm rad/m^{2}}$ and a variance of $20 \ {\rm rad/m^{2}}$. We can now assign to every source in the catalogue a value for $\phi_{0}$ and $\psi$. The full list of entries of our catalogue era reported in table~\ref{tab:PS_cat}. 
The final differential polarized source counts is reported in figure~\ref{fig:pol_sc}. The full catalogue is available in the \href{https://drive.google.com/drive/folders/1bbm7ExfkA1-jT-o8yme5DaeXMWOrjBTd?usp=sharing}{CRC repository}. We recall that the catalogue is based on simplified assumptions. 
Nevertheless, polarization leakage is still a not well explored topic and thus we believe that this polarized source catalogue could be of general interest for the IM community.

\begin{table}\label{tab:PS_cat}
\centering
\caption{The $1.4$~GHz point source catalogue catalogue format for every radio source. The total number of sources is $\sim 408.8$ million.}
\begin{tabular}{c|c|c}
\hline
ra  &   right ascension in degrees\\
dec & declination in degrees \\
$\alpha$ & source spectral index \\
$S_{1.4\:{\rm GHz}}$ & source flux density at $1.4$~GHz in mJy\\
$|P|$ & source polarized flux density at $1.4$~GHz in mJy\\
$\phi_0$ & the intrinsic polarisation angle in degrees\\
$\psi$ & the rotation measure given in ${\rm rad}/m^2$\\
\hline
\end{tabular}
\end{table}

\section{Map deconvolution}\label{app:resmooth}
In section~\ref{sec:resmooth}, we have discussed the performance of the foreground cleaning on maps that have been deconvolved to a common resolution. Here we present in more details some of the steps of the adopted procedure.

\medskip \noindent {\it Apodization.}
Since we are interested only in how well we reconstruct the signal and not in the shape of the signal itself, throughout the paper we have computed the angular power spectrum of the maps using simply equation~\ref{eq:cl} and correcting for $f_{\rm sky}$ (see also discussion in section~\ref{sec:fg_diag}). Nevertheless, to re-smooth a map using equation~\ref{eq:alm_res}, having well behaving $a_{\ell m}$ becomes important. To this purpose we do not compute them directly using the original sharp mask but we use an {\rm apodized} mask. Apodization is a standard procedure for example in CMB studies, assuring that the mask does not sharply pass from one to zero but that there is a smooth transition at the borders. Apodized masks alleviate the {\it ringing} in the power spectrum, and thus in the $a_{\ell m}$. To apodize, we smooth the mask with a Gaussian with a FWHM of a few degrees, replacing the zeros with negative numbers before the smoothing to obtain a better behaviour around the edges of the mask. We choose this negative value ensuring that, in the sky patch defined by the initial mask, there are no negative values. The final apodized mask is obtained just by multiplying the smoothed one and the original mask.\\

\begin{figure}
\includegraphics[width=8cm]{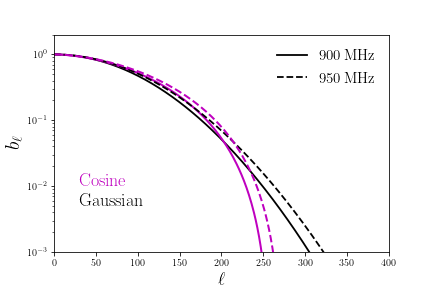}
\caption{Spherical harmonic coefficients of the beam decomposition $b_\ell$ (see equation~\ref{eq:b_l}) for the cosine beam (magenta) and the Gaussian beam (black). The lower frequency of L-band 900 MHz is plotted with a solid line while $950~\mathsf{MHz}$ is plotted with a dashed line.}
 \label{fig:bl}
 \end{figure}

\medskip \noindent {\it Beam $b_\ell$ ratio.}
The other important ingredient for equation~\ref{eq:alm_res} are the spherical harmonics coefficient of the beam decomposition on the sphere, the $b_\ell$ of equation~\ref{eq:bl}. As discussed in section~\ref{sec:beamconv}, for a symmetric beam these coefficients are only function of $\ell$ and are real.
We report in figure~\ref{fig:bl} the shape of the $b_\ell$ for two different frequencies and comparing the cosine and the Gaussian beams. Due to the shape of the cosine main lobe (see figure~\ref{fig:beam_models}), the $b_\ell$ drops faster after $\ell \sim 200$. 

\begin{figure}
\includegraphics[width=8cm]{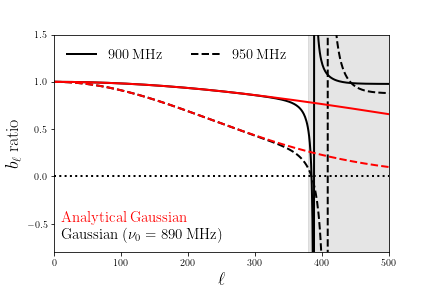}
\includegraphics[width=8cm]{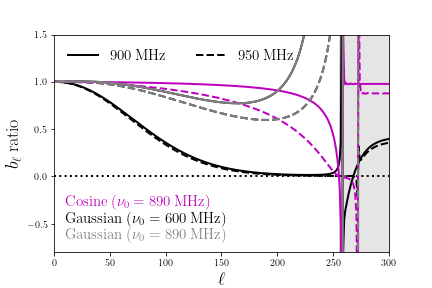}
\caption{{\it Top panel:} The ratio between the $b_\ell$ for a Gaussian beam at $900~\mathsf{MHz}$ (solid black) or at $950~\mathsf{MHz}$ (dashed black) with the lower frequency Gaussian beam at $\nu_0=890~\mathsf{MHz}$. After $\ell \sim 380$ the ratio is not well behaved so we artificially set to zero the smaller scales. This approximated procedure is compared with its the analytical solution for both $900~\mathsf{MHz}$ (solid red) and at 950 MHz (dashed red). {\it Bottom panel:} The ratio between the $b_\ell$ for a cosine beam at 900 MHz (solid magenta) or at 950 MHz (dashed magenta) with the lower frequency cosine beam at $\nu_0=890$~MHz. In this case after $\ell \sim 240$ the ratio is not well behaved so we consider zero all the smaller scales. We report also the behaviour of the ratio when we deconvolve with a cosine beam but re-convolve with a Gaussian. If we use the same$\nu_0$ as before, i.e. $890~\mathsf{MHz}$ the ratio start a fast growth around $\ell\sim 200$ (grey lines), while to obtain a reasonable value for the ratio we are obliged to throw away a lot of information using a very large beam, i.e. a very low $\nu_0$ (black lines).}
 \label{fig:bl_ratio}
 \end{figure}

In section~\ref{sec:resmooth} we discussed different cases for re-smoothing. In the first case, we have used a Gaussian both to de-convolve and re-convolve, i.e. the ratio of equation~\ref{eq:alm_res}. In harmonic space a Gaussian function remains Gaussian, so we have an easy analytical form for the ratio: it corresponds to the $b_\ell$ of a Gaussian beam with $\Delta \theta^2=\Delta \theta(\nu_0)^2-\Delta \theta (\nu)^2$. Note that, as discussed in \citet{anderson2018}, it is better to choose $\nu_0$ to be slightly lower than the minimum frequency used (we use $890~\mathsf{MHz}$).
For a better comparison, instead of using the analytical form we compute numerically the ratio between the two $b_\ell$. The finite machine precision inevitably introduces spurious behaviors at small scales ($\ell \sim 370$ for our frequency range, see top panel of figure~\ref{fig:bl_ratio}). To avoid the effect of these artifacts, in case of numerical ratio, one can perform a cut in $\ell$ and artificially put at zero the ratio for smaller scales.
We compare the numerical and the analytical case in the top panel of figure~\ref{fig:bl_ratio}, showing the loss of information imposed by the $\ell_{\rm cut}$.

For the re-smoothing case where we assume that the beam could be described with a {\it Cosine} model, the analytical ratio is not straightforward to derive so we always revert to the numerical ratio. In the bottom panel of figure~\ref{fig:bl_ratio}, we show how the ratio behaves with this beam model. Due to the differences in the $b_\ell$ seen in figure~\ref{fig:bl}, the effect of finite machine precision enters at larger scale so we perform the cut at $\ell\sim 250$.

One could also try to deconvolve the maps form the effect of the {\it Cosine} beam and then use a {\it Gaussian} re-smoothing to alleviate the effect of the sidelobes.
In the bottom panel of figure~\ref{fig:bl_ratio} we report also this case, noting however that a very large Gaussian beam is required to avoid spurious numerical behaviour of the ratio. If on the one hand this could be of help in suppressing artifacts in the data, on the other hand we are forced to throw away more information. We have thus decided to use the cosine beam also for re-smoothing.

\label{lastpage}
\end{document}